\documentclass[11pt,a4paper]{article}
\pdfoutput=1
\usepackage{jheppub}
\usepackage{graphicx,amssymb,amsmath,amsfonts}
\usepackage[english]{babel}
\usepackage{slashed}
\usepackage{multirow}

\def\mL{\mathcal{L}}
\def\mS{\mathcal{S}}
\def\mW{\mathcal{W}}
\def\mWb{\overline{\mathcal{W}}}
\def\mK{\mathcal{K}}

\def\mM{\mathcal{M}}
\def\mO{\mathcal{O}}
\def\mN{\mathcal{N}}
\def\mD{\mathcal{D}}
\def\mM{\mathcal{M}}
\def\mS{\mathcal{S}}
\def\mQ{\mathcal{Q}}
\def\mR{\mathcal{R}}
\def\mH{\mathcal{H}}

\def\zb{\bar{z}}
\def\taub{\bar{\tau}}
\def\ubar{\bar{u}}
\def\vbar{\bar{v}}
\def\ebar{\bar{e}}
\def\Ebar{\bar{E}}

\def\xb{\bar{x}}

\def\ab{\bar{a}}
\def\bbar{\bar{b}}
\def\mb{\bar{m}}

\def\fb{\bar{f}}
\def\Zb{\bar{Z}}
\def\lamb{\bar{\lambda}}
\def\chib{\bar{\chi}}
\def\Wb{\overline{W}}

\def\pa{\partial}

\def\Kahler{K{\"a}hler }

\def\g5{\gamma_5}
\def\gmu{\gamma^{\mu}}
\def\gmd{\gamma_{\mu}}

\def\lam{\lambda}

\def\zetat{\tilde{\zeta}}

\def\sigmab{\bar{\sigma}}
\def\Ft{\tilde{F}}

\def\mt{\tilde{m}}

\def\Et{\tilde{E}}
\def\Ht{\tilde{H}}
\def\Ft{\tilde{F}}
\def\Yt{\tilde{Y}}

\def\b[#1]{\bold{#1}}
\def\bb[#1]{\overline{\bold{#1}}}

\def\bs[#1,#2]{\bold{#1}_{#2}}
\def\bbs[#1,#2]{\overline{\bold{#1}}_{#2}}

\def\s2{\sigma_2}

\def\ep{\epsilon}

\def\D{\Delta}

\makeatother

\title{Holography for $\mN=1^*$ on $S^4$ and Supergravity}
\author{Uri Kol}

\affiliation{Michigan Center for Theoretical Physics, \\
Randall Laboratory of Physics, Department of Physics,\\
University of Michigan, Ann Arbor, MI 48109, USA}

\emailAdd{urikol@umich.edu}

\abstract{
We study the $SO(3)$ sector of the $\mN=1^*$ mass deformation of $\mN=4$ super Yang-Mills on $S^4$.
The gravity dual of this sector is $\mN=2$ supergravity coupled to two hypermultiplets.
The scalar fields in the hypermultiplets span a quaternionic-\Kahler manifold that is described by the coset $G_{2,2}/SU(2)\times SU(2)$.

We use the $\mN=2$ supergravity dual to study field configurations in the bulk that feature analytical solutions, and compute the corresponding $S^4$ free energy using the procedure of holographic renormalization.
We find that the free energy of these configurations is quadratic in the mass and show that it is devoid of unphysical ambiguities, hence providing an analytical prediction for the $\mN=1$ four-sphere partition function at large 't Hooft coupling in the planar limit.
}

\begin{document}
\maketitle

\section{Introduction and summary of the results}

Recently there has been a growing interest in the study of supersymmetric gauge theories on curved manifold. In his seminal work \cite{Pestun:2007rz}, Pestun used localization techniques to evaluate the partition function of various $\mN=2$ supersymmetric gauge theories on $S^4$. Following Pestun's work, localization techniques have been extensively used to study field theories at finite and strong coupling. However, since localization requires the existence of at least $\mN=2$ supersymmetries \cite{Knodel:2014xea}, very little is known about theories with a smaller amount of supersymmetry.
A great progress was made by Festuccia and Seiberg \cite{Festuccia:2011ws}, who have studied the coupling of supersymmetric field theories to curved manifolds in three and four spacetime dimensions using rigid supergravity. Their work shed light on the kinematics of $\mN=1$ theories on curved manifolds, but did not address their dynamics.
Exact results for $\mN=1$ theories on $S^4$ are therefore still mostly out of reach.

The motivation behind the work presented in \cite{Bobev:2016nua} was to change this situation and derive exact results for $\mN=1$ theories. The authors of \cite{Bobev:2016nua} used the gauge/gravity correspondence to study the $\mN=1^*$ theory on $S^4$ and at strong coupling, using its gravity dual.
Let us briefly review the $\mN=1^*$ mass deformation of $\mN=4$ super Yang-Mills.
In Lorentzian signature it is described by the following superpotential (and its complex conjugate)
\begin{equation}\label{N=1superpotential}
\begin{aligned}
\mW &= \sqrt{2} g_{YM} f^{abc} Z_1^a Z_2^b Z_3^c +\frac{1}{2} \sum_{i=1}^3 m_{ij} Z_i ^a Z_j^a \\
\mWb &= \sqrt{2} g_{YM} f^{abc} \Zb_1^a \Zb_2^b \Zb_3^c +\frac{1}{2} \sum_{i=1}^3 \mb_{ij} \Zb_i ^a \Zb_j^a
\end{aligned}
\end{equation}
where $g_{YM}$ is the Yang-Mills coupling and $f^{abc}$ are the structure constants of the gauge group normalized in a way independent of the $g_{YM}$.
However, in order to couple the theory to the sphere one has to perform Euclidean continuation.
Fields that are related by complex conjugation in the Lorentzian theory, are independent in Euclidean signature \cite{Freedman:2013ryh,Bobev:2013cja}.
In the Euclidean theory, the mass parameters $m_{ij}$ and $\mb_{ij}$ are therefore also independent.

Before calculating anything, we have to ask whether a supersymmetric observable, like the partition function, computed on a curved manifold is free of renormalization scheme ambiguities.
Indeed, in the case of superconformal $\mN=1$ theories in four dimensions, for example, it was shown in \cite{Gerchkovitz:2014gta} that the partition function on $S^4$, seen as a function of exactly marginal couplings, is completely scheme dependent. The situation is different for $\mN=2$ superconformal theories where the $S^4$ partition function can be expressed in terms of the K\"{a}hler potential of the Zamolodchikov metric and thus contains physically interesting information.
The analysis of \cite{Gerchkovitz:2014gta} can be extended to massive theories. In \cite{Bobev:2016nua}, it was shown that the ambiguities in the free energy of the $\mN=1^*$ theory are of the following form
\begin{equation}\label{ambiguities}
F_{\mN=1^*} \rightarrow
F_{\mN=1^*} + f_1 (\tau) +\fb_1(\taub)+a^2 \sum_{i,j=1}^3 (m_{ij} \mb_{ij}) f_2(\tau,\taub)
\end{equation}
where $f_1,\fb_1$ and $f_2$ are arbitrary functions of the complexified gauge coupling
\begin{equation}
\tau = \frac{\theta}{2\pi} +i \frac{4 \pi}{g_{YM}^2}
\end{equation}
and $a$ is the radius of the four-sphere.
The holomorphic structure of the UV ambiguities $f_1,\fb_1$ is a result of the extended supersymmetry of the UV theory, which is $\mN=4$ SYM.
$\mN=4$ supersymmetry can be regarded as a particular case of $\mN=2$ supersymmetry and, as was explained in \cite{Gerchkovitz:2014gta}, the ambiguities in the sphere partition function are then understood as \Kahler ambiguities. In the case where the UV theory preserves only $\mN=1$ supersymmetry, the structure of ambiguities is encoded in a single non-holomorphic function $f_1(\tau,\taub)$ and the ambiguities are not physical.
We would like to emphasize that, unlike the physically understood $f_1,\fb_1$ ambiguities, the second ambiguity $f_2$ of the massive theory is a real unphysical ambiguity. Physical observables therefore cannot be subject to ambiguities of the form represented by $f_2(\tau,\taub)$.

The gravity dual of $\mN=4$ SYM is type IIB supergravity on $AdS_5 \times S^5$ \cite{Maldacena:1997re}.
The gravitational description can be simplified by using $\mN=8$ gauged supergravity in five dimensions, which is a consistent truncation of type IIB supergravity on $S^5$.
The mass deformations encoded in the superpotential \eqref{N=1superpotential} correspond to turning on scalar fields in the $\mN=8$ supergravity theory.
One can therefore study the massive theory using five-dimensional domain-wall solutions in that gravity side that involve scalar fields coupled to the metric.
This approach was taken in the past to study holographic RG flows in flat spacetime.
However, to study the theory on a curved background one has to include additional couplings due to the curvature, as was explored in \cite{Bobev:2016nua}.

In this paper we extend the analysis of \cite{Bobev:2016nua}. Our main motivation is to derive analytical expressions for the $S^4$ free energy.
We will be mainly interested in the equal mass case
\begin{equation}\label{equalMassCase}
m_{ij} = m \delta_{ij}, \qquad \mb_{ij} = \mb \delta_{ij}
\end{equation}
with $m\neq \mb$. This theory preserves, in addition to the $U(1)_R$ symmetry, an $SO(3)$ global symmetry inside the original $SU(4)$ symmetry of $\mN=4$ SYM.
On the gravity side, the $SO(3)$ sector is described by an $\mN=2$ truncation of the maximally supersymmetric $\mN=8$ supergravity theory \cite{Pilch:2000fu}. In addition to the supergravity multiplet, it also contains two hypermultiplets.
The scalars in the hypermultiplets span a manifold that is described by the coset
\begin{equation}\label{G2coset}
\frac{G_{2,2}}{SU(2)\times SU(2)}
\end{equation}
where $G_{2,2}$ is the non-compact form of the exceptional Lie group $G_2$. This coset is known to describe a quaternionic-\Kahler manifold \cite{Ferrara:1989ik,Bodner:1989cg,Castellani:1983tb}.
In general, quaternionic-\Kahler manifolds are described using a triplet of \emph{prepotentials} $P^r$, where $r=1,2,3$ is an index in the adjoint representation of $SU(2)_R$ (the R-symmetry group of $\mN=2$ supergravity).
When the matter sector includes only hypermultiplets, like in the case under consideration, the theory can be described using a superpotential
\begin{equation}
W=\sqrt{\frac{2}{3}P^r P^r}
\end{equation}
(not to be confused with the superpotential of the field theory $\mW$).
We study the quaternionic-\Kahler manifold \eqref{G2coset} and derive a superpotential for it.

Finally, we focus on the following mass configuration
\begin{equation}\label{configuration}
m\neq 0 , \qquad \mb=0
\end{equation}
namely, we set the masses of the anti-chiral multiplets to zero, while keeping the masses of the chiral multiplets non-zero. In Lorentzian signature it is not possible, but as explained above, Euclidean theories allow for these configurations.
There are two motivations to look at the configuration \eqref{configuration}. First, as evident from \eqref{ambiguities}, when $\mb=0$ the unphysical ambiguity vanishes and the sphere partition function is well-defined.
Second, as we will show later, there is a field configuration in the bulk that correspond to \eqref{configuration} and for which the scalar kinetic term vanishes. In such a case the stress-tensor vanishes as well, and the scalars do not back-react on the metric. By the Einstein equations, the metric is then simply given by the hyperbolic space $\mathbb{H}^5$
\begin{equation}
ds^2 = dr^2 +(\sinh r)^2 ds^2_{S^4}
\end{equation}
The conformal symmetry is still broken because of the non-trivial profile of the scalars in the bulk.
In Lorentzian signature, a situation where the metric is $AdS_5$ but the matter fields break the $SO(4,2)$ symmetry would be impossible, because any complex scalar with a non-trivial profile in the $AdS_5$ directions produces a non-vanishing stress tensor.
Similar configurations were found in \cite{Freedman:2013ryh} for the ABJM theory on $S^3$.

The configuration described above features an analytical solution in the bulk, which we use to compute the free energy using the procedure of holographic renormalization. The result is
\begin{equation}\label{FreeEnergy}
F_{S^4} = F_0 -  \frac{15}{128} N^2 (ma)^2
\end{equation}
$F_0$ is the free energy of $\mN=4$ SYM on $S^4$
\begin{equation}
F_0 = -\frac{N^2}{2} \log \lambda
\end{equation}
where $\lambda$ is the 't Hooft coupling $\lambda = g_{\text{YM}}^2 N$ \cite{Russo:2012ay,Buchel:2013id,Russo:2013sba,Crossley:2014oea}.
The free energy \eqref{FreeEnergy} is well-defined and devoid of unphysical ambiguities.
The result \eqref{FreeEnergy} provides an analytical prediction for the free energy of the theory at strong coupling.
This is our main result.

The paper is organized as follows.
In section \ref{FieldTheory} we provide a brief review of the $\mN=1^*$ theory.
In section \ref{supergravity} we review the structure of $\mN=2$ supergravity and derive the equations of motion for domain-wall solutions with $S^4$ boundary.
As a warmup exercise, we describe the universal hypermultiplet, as part of the Leigh-Strassler flow, in section \ref{universalHyperSection}.
The reader who is familiar with $\mN=2$ supergravity can skip directly to section \ref{G2section}, where we describe the $G_{2,2}$ coset model and derive the superpotenial.
In section \ref{SolutionsSec} we discuss several solutions of the coset model, including the ones with no back-reaction.
In section \ref{FreeEnergySec} we calculate the free energy using the procedure of holographic renormalization.
We end with conclusions and future directions \ref{conclusions}.
Few appendices include more information about the theory and the calculation.

\section{Field theory}\label{FieldTheory}

We start by reviewing the $\mN=1^*$ mass deformation of the $\mN=4$ supersymmetric Yang-Mills theory in flat space and on the four-sphere \cite{Bobev:2016nua}.
$\mN=4$ super Yang-Mills can be written in an $\mN=1$ language using three chiral multiplets and a superpotential given by \eqref{N=1superpotential} with the masses set to zero.
In this form only an $\mN=1$ supersymmetry is manifest, but the full $\mN=4$ supersymmetry is still preserved.
When the masses in \eqref{N=1superpotential} are non-zero the $\mN=4$ supersymmetry is broken down to $\mN=1$ supersymmetry.
The kinetic term is given by
\begin{equation}\label{kineticTerm}
\begin{aligned}
\mL_{\text{kinetic}} &=
\frac{1}{4g_{YM}^2} \left( F^a_{\mu\nu} \right)^2
+\frac{\theta}{16 \pi ^2} \ep^{\mu\nu\rho\sigma} F^a_{\mu\nu}F^a_{\rho\sigma}
-\lamb^a\,^T \s2 \sigmab^{\mu}D_{\mu} \lam^a \\
&+D^{\mu}\Zb^a_i D_{\mu}Z^a_i
-\chib^a_i\,^T \s2 \sigmab^{\mu}D_{\mu} \chi^a_i 
\end{aligned}
\end{equation}
$F^a_{\mu\nu}$ is the gauge field strength, $\lam^a_{\alpha}$ and $\lamb^{a\dot{\alpha}}$ are the left-handed and right-handed components of the gauginos, $Z^a_i$ are the bottom components of the chiral multiplets and $\Zb^a_i$ are their conjugates, and $\chi^a_{i\alpha}$ and $\chib^{a\dot{\alpha}}_i$ are the left-handed and right-handed components of the fermions in the chiral multiplets. Fields that in Lorentzian signature are related by complex conjugation, are independent in Euclidean signature.

The interaction Lagrangian in flat space is given by
\begin{equation}
\begin{aligned}
\mL_{\text{interaction}} &=
\Wb^a_i W^a_i +\frac{1}{2}\left( \chi^a_i\,^T \s2 \chi^b_j \right) W^{ab}_{ij}
+\frac{1}{2}\left( \chib^a_i\,^T \s2 \chib^b_j \right) \Wb^{ab}_{ij}
\end{aligned}
\end{equation}
where subscripts on $W,\Wb$ represent derivatives with respect to the scalars.
In order to couple the $\mN=1^*$ theory to the four-sphere while preserving supersymmetry we need to include also the following terms (as well covariantizing the derivatives in \eqref{kineticTerm})
\begin{equation}
\begin{aligned}
\mL_{S^4-\text{interaction}} &=\frac{2}{a^2}\Zb^a_i Z^a_i \mp\frac{i}{a}\left(  3W-W^a_iZ^a_i \right)
\mp\frac{i}{a}\left(  3\Wb-\Wb^a_i \Zb^a_i \right)
\end{aligned}
\end{equation}
The first term is nothing but the conformal coupling to the sphere while the other terms are needed to preserve supersymmetry on the sphere \cite{Festuccia:2011ws}.

With the superpotential \eqref{N=1superpotential} the resulting Lagrangian for $\mN=1^*$ on $S^4$ is given by
\begin{equation}
\mL_{\mN=1^*}^{S^4}=
\mL_{\text{kinetic}}
+\mL_2
+\mL_{\text{Yukawa}}
+\mL_3
+\mL_4
\end{equation}
The quadratic interaction term is
\begin{equation}\label{quadraticL}
\begin{aligned}
\mL_2 &= \mb_{ij} m_{ik} \Zb^a_j Z^a_k
-\frac{1}{2}m_{ij} \left(  \chi^a_i \, ^T \s2 \chi^a_j  \right)
-\frac{1}{2}\mb_{ij} \left(  \chib^a_i \, ^T \s2 \chib^a_j  \right) \\
&+\frac{2}{a^2}\Zb^a_i Z^a_i
\pm \frac{i}{2a}\left(  m_{ij} Z^a_i Z^a_j + \mb_{ij} \Zb^a_i \Zb^a_j   \right)
\end{aligned}
\end{equation}
The Yukawa and cubic interaction terms are respectively given by
\begin{equation}
\begin{aligned}
\mL_{\text{Yukawa}} &=
\sqrt{2} g_{YM}f^{abc}\left[
\left( \lam^a\,^T \s2 \chi^b_i \right) \Zb^c_i
+\left( \lamb^a\,^T \s2 \chib^b_i \right) Z^c_i
\right. \\ &  \left. 
+\frac{1}{2}\ep_{ijk}\left( \chi^a_i \,^T \s2 \chi^b_j  \right) Z^c_k
+\frac{1}{2}\ep_{ijk}\left( \chib^a_i \,^T \s2 \chib^b_j  \right) \Zb^c_k
\right]\\
\mL_3 &= -\frac{g_{YM}}{\sqrt{2}} f^{abc} \ep_{ijk} \left(
\mb_{il} \Zb^a_l Z^b_j Z^c_k
+m_{il} Z^a_l \Zb^b_j \Zb^c_k
\right)
\end{aligned}
\end{equation}
and the quartic interaction term is given by
\begin{equation}
\mL_{4} =
\frac{g_{YM}^2}{2}f^{abc}f^{ade}
\left(
-\Zb^b_i Z^c_i \Zb^d_j Z^e_j
+2 \Zb^b_j \Zb^c_i Z^d_j Z^e_i
\right)
\end{equation}
For more details see \cite{Bobev:2016nua}.

In this paper we will focus on the equal mass case \eqref{equalMassCase}.
In this case, the first term in \eqref{quadraticL} is proportional to the Konishi operator $K=|Z_1|^2+|Z_2|^2+|Z_3|^2$, which is invisible in the supergravity limit.
We are therefore left with the following four massive operators in the Lagrangian
\begin{equation}\label{SO(3)Lagrangian}
\begin{aligned}
\mL_{SO(3)} & = 
-\frac{1}{2} m \left( \chi^a\,^T \s2 \chi^a  \right)
-\frac{1}{2} \mb \left( \chib^a\,^T \s2 \chib^a  \right)
\pm \frac{i}{2a} 
\left(  m Z^2 +\mb\Zb^2  \right)
\end{aligned}
\end{equation}
(the cubic interaction terms $\mL_3$ are in the same R-symmetry representations as the fermion bilinears, and are therefore indistinguishable from them). 
In addition to the four massive operators, the Lagrangian includes the gauge kinetic term and the $\theta$-term.
Finally, we also have to take into account left-handed and right-handed gaugino bilinears, that can possibly condense.
In total, the spectrum of the $SO(3)$ sector includes eight scalar operators.
We therefore expect to have eight dual scalar fields in the bulk.

\section{$\mN=2$ Supergravity}\label{supergravity}

In this section we review the general structure of $\mN=2$ supergravity in five space-time dimensions (see \cite{Ceresole:2000jd,Ceresole:2001wi,deWit:1991nm,deWit:1992cr,Andrianopoli:1996vr,Andrianopoli:1996cm} for reference). The theory contains the supergravity multiplet and can be coupled to matter fields.
The pure supergravity multiplet
\begin{equation}
\left(   e^a_{\mu}, \psi^{\alpha i}_{\mu} , A_{\mu}  \right)
\end{equation}
contains the graviton $e^a_{\mu}$, two gravitini $\psi^{\alpha i}_{\mu} $ and a vector field $A_{\mu}$ (the graviphoton).
The supergravity multiplet can be coupled to vector, tensor and hyper multiplets.
Each vector multiplet contains one gauge field, two gauginos and one real scalar
\begin{equation}
\left(  A_{\mu} , \lambda_i , \phi \right)
\end{equation}
Each hypermultiplet contains two hyperinos and four real scalars
\begin{equation}
\left( \zeta^A , q^X \right)
\end{equation}
We will not consider here tensor multiplets.

We start by describing the general structure of one supergravity multiplet coupled to $n_V$ vector multiplets and $n_H$ hypermultiplets \cite{Ceresole:2001wi}.
The scalar manifold in this case is a direct product of a "very special manifold" \cite{deWit:1991nm,deWit:1992cr} and a quaternionic K\"{a}hler manifold
\begin{equation}
\mM = \mS(n_V) \times \mQ (n_H)
\end{equation}
The $\mS$ manifold is the $n_V$-dimensional target space of the $\phi^x$ scalars and $x=1,\dots,n_V$ are the curved indices labeling the coordinates on $\mS$.
The $\mQ$ manifold is the $4n_H$-dimensional target space of the $q^X$ scalars and $X=1,\dots,4n_H$ are the curved indices labeling the coordinates on $\mQ$.
The holonomy group of the manifold $\mQ$ is a direct product of $USp(2)\simeq SU(2)$ and some subgroup of the symplectic group in $2n_H$ dimensions
\begin{equation}
G\equiv\mathcal{H}ol(\mQ) = SU(2) \times USp(2n_H)   \subset SU(2) \times Sp(2n_H,\mathbb{R})  
\end{equation}
The $SU(2)$ factor is the R-symmetry group and the index $i=1,2$ corresponds to it fundamental representation.
The index $A=1,\dots,2n_H$ correspond to the fundamental representation of $USp(2n_H)$.

The gauging of matter fields coupled to $\mN=2$ supergravity theory is achieved by identifying the gauge group $K$ as a subgroup of the isometries $G$ of the product space $\mM$. Two main cases are known in the literature (see \cite{Andrianopoli:1996vr,Andrianopoli:1996cm} for reviews):
\begin{enumerate}
\item $K$ is non-Abelian
\item $K = U(1)^{n_V+1}$
\end{enumerate}
In the first case, supersymmetry requires $K$ to be a subgroup of the full $\mM$. In the Abelian case, the $\mS$ manifold is not required to have any gauged isometries.
The action of the gauge group $K$ on the scalar manifold $\mM$ is
\begin{equation}
\begin{aligned}
q^X &\rightarrow q^X + \ep^I K^X_I(q) \\
\phi^x &\rightarrow \phi^x + \ep^I K^x_I(\phi)
\end{aligned}
\end{equation}
for infinitesimal parameter $\ep_I$.
The index $I=0,\dots,n_V$ runs over the gauge fields (one graviphoton plus $n_V$ gauge fields of the vector multiplets).
$K^X_I(q)$ are the Killing vectors of the gauged isometries on the quaternionic scalar manifold and $K^x_I(\phi)$ are those of the very special manifold.

\subsection{The $\mS$ manifold}

The scalars of the vector multiplets can be described by a hypersurface in an $(n+1)$-dimensional space
\begin{equation}
C_{IJK}h^I(\phi)h^J(\phi)h^K(\phi)=1
\end{equation}
The real coefficients $C_{IJK}$ determine the metric of a "very special geometry" \cite{deWit:1991nm,deWit:1992cr} 
 \begin{equation}\label{vectorSurface}
\begin{aligned}
a_{IJ} \equiv& -2 C_{IJK} h^K +3C_{IKL}C_{JMN} h^K h^L h^M h^N = h_I h_J + h_{xI} h^x_J, \\
g_{xy} \equiv&  h^I_x h^J_y a_{IJ}, \qquad h_I=C_{IJK}h^I h^K, \qquad h^I_x \equiv - \sqrt{\frac{3}{2}} \pa_x h^I(\phi)
\end{aligned}
\end{equation}

\subsection{The $\mQ$ manifold}

The quaternionic K\"{a}hler geometry is determined by $4n_H$-beins $f^{iA}_X$. The $SU(2)$ index $i=1,2$ is raised and lowered by the $\ep$ symbol. The $Sp(2n_H)$ index $A=1,\dots,2n_H$ is raised and lowered by the symplectic matrix $C_{AB}$ (see appendix \ref{su2structure}).

The metric on the hyperscalar manifold is given by
\begin{equation}
g_{XY} = f^{iA}_X f^{jB}_Y \ep_{ij}C_{AB} = f^{iA}_Xf_{YiA}
\end{equation}
This implies that the vielbeins satisfy
\begin{equation}
f^X_{iA}f^{iA}_Y = \delta^X_Y, \qquad f^X_{iA}f^{jB}_X = \delta^j_i\delta^B_A
\end{equation}
and they are also covariantly constant 
\begin{equation}
\pa_X f^{iA}_Y-\Gamma^Z_{XY}f^{iA}_Z+ f^{iB}_Y \omega_{XB}\,^A +\omega_{Xk}\,^i f^{kA}_Y =0
\end{equation}
$\Gamma^Z_{XY}$ is the Levi-Civita connection on the hyperscalar manifold, $\omega_{XB}\,^A$ is the $Sp(2n_H)$ connection and $\omega_{Xi}\,^j$ is the $SU(2)$ spin connection.
The $SU(2)$ curvature is
\begin{equation}
\mR_{XYij} = f_{XA(i} f^A_{j)Y}
\end{equation}
The $SU(2)$ curvature can be expressed in terms of the $SU(2)$ spin connection
\begin{equation}
\mR_{XYi}\,\!^j =  2 \pa_{[X}\omega_{Y]i}\,\!^j
-2 \omega_{[Xi}\,\!^k\omega_{Y]k}\,\!^j
\end{equation}
The $SU(2)$ curvature can be decomposed in terms of $SU(2)$ triplets
\begin{equation}
\mR_{XYi}\,\!^j 
=i \mR^r_{XY} \left( \tau^r \right)_i \,\!^j
\end{equation}
where $r=1,2,3$ and $ \left( \tau^r \right)_i \,\!^j$ are the three Pauli matrices (see appendix \ref{su2structure}).
The triplet of curvatures satisfy the following identity
\begin{equation}\label{Ridentity}
\mR^r_{XY} \mR ^{s YZ} = -\frac{1}{4} \delta^{rs} \delta_X \,\! ^Z  + \frac{1}{2} \ep^{rst} \mR ^t _X \,\! ^Z
\end{equation}
In differential form the curvature triplets are expressed in terms of spin connection triplets as
\begin{equation}\label{curvature}
\mR^r = d \omega^r -\ep^{rst}\omega^s \omega ^t
\end{equation}
The prepotentials associated with the Killing vectors are given by
\begin{equation}\label{prepotentialFormula}
P^r_I = \frac{1}{2 n_H} \mR^{r XY} D_X K_{IY}
\end{equation}
and the inverse relation is
\begin{equation}
K^X_I = - \frac{4}{3} \mR^{r XY} D_Y P^r_{I}
\end{equation}
For more details see \cite{Ceresole:2001wi}.

\subsection{The Lagrangian}

The bosonic part of the Lagrangian, of an $\mN=2$ supergravity coupled to $n_V$ vector multiplets and $n_H$ hypermultiplets, in Lorentzian signature is
\begin{equation}
\begin{aligned}
\mL= & -\frac{1}{2} R  -\frac{1}{4} a_{IJ} F_{\mu\nu}^I F^{J \mu\nu}
-\frac{1}{2} g_{XY} \mD_{\mu}q^X \mD^{\mu}q^Y
-\frac{1}{2} g_{xy} \mD_{\mu}\phi^x \mD^{\mu}\phi^y
\\
&+\frac{1}{6\sqrt{6}} C_{IJK} \ep^{\mu\nu\rho\sigma\tau}F^I_{\mu\nu}F^J_{\rho\sigma}A^K_{\tau}
-g^2 V
\end{aligned}\label{Lagrangian}
\end{equation}
where we are using the mostly minus signature and the gauge-covariant derivatives are
\begin{equation}
\begin{aligned}
\mD _{\mu} q^X &= \pa_{\mu}q^X+ g A^I_{\mu}K^X_I(q) \\
\mD _{\mu} \phi^x &= \pa_{\mu}\phi^x+ g A^I_{\mu}K^x_I(\phi) 
\end{aligned}
\end{equation}
With these notations the coupling $g$ is related to the AdS radius $L$ via
\begin{equation}
g=\frac{1}{L}
\end{equation}
From now on we set the AdS radius to $L=1$.

\subsection{Supersymmetry transformations}

The bosonic part of the supersymmetry transformations of the fermions (with vanishing vectors) are
 \begin{equation}
\begin{aligned}
\delta \psi_{\mu i} =& D_{\mu}(\omega) \ep_i  - \omega_{\mu i}\, ^j\ep_j +i \frac{1}{\sqrt{6}}g\gamma_{\mu}P_{ij}\ep^j \\
\delta \lambda_i^x =& -i \frac{1}{2} (\slashed \pa \phi^x) \ep_i +g P^x_{ij}\ep ^j \\
\delta \zeta ^A =& -i \frac{1}{2} f^A_{iX}(\slashed \pa q^X) \ep^i +g\ep^i \mN ^A _i
\end{aligned}\label{susyVar}
\end{equation}
where $\omega_{\mu i}\,^j = (\pa_{\mu}q^X) \omega_{Xi}\,^j$ and $D_{\mu}(\omega)=\pa_{\mu} + \frac{1}{4} \gamma_{ab}\omega_{\mu}^{ab} $ ($\omega_{\mu}^{ab}$ is the spacetime connection).
The two spinors $\ep_i$ obey the symplectic Majorana condition (see appendix \ref{Clifford5D} for more details on the gamma matrices in five dimensions)
\begin{equation}
\ep_2 = \g5 \ep_1^*
\end{equation}

$\mN^A_i$ is a function of the Killing vectors associated with the gauged isometries
\begin{equation}
\mN^A_i = \frac{\sqrt{6}}{4}f^A_{iX} K^X, \qquad K^X=h^I(\phi)K^X_I(q)
\end{equation}
$P_{ij}$ is a function of the prepotentials associated with the gauged isometries $P^r_I$.
The dependence is as follows: first, $P_{ij}$ can be decomposed in terms of $SU(2)$ triplets $P^r$ (see appendix \ref{su2structure})
\begin{equation}
P_{ij} = i P^r (\tau^r)_{ij}
\end{equation}
which are, in turn, related to the prepotentials
\begin{equation}
P^r = h^I(\phi) P^r_I(q)
\end{equation}
In addition, we define
\begin{equation}
P^r_x=-\sqrt{\frac{3}{2}}\pa_xP^r, \qquad P^x_{ij}= -\sqrt{\frac{3}{2}}\pa^x P_{ij}
\end{equation}

\subsection{The scalar potential and a Bogomolnyi form}

The scalar potential is given by the following expression
\begin{equation}
V=-4 P^r P^r +2 P^r_x P^r_y g^{xy} +2\mN_{iA} \mN^{iA}
\label{potential}
\end{equation}
In some cases the scalar potential can be brought to the Bogomolnyi form which is described by an $\mN=1$ superpotential.
To show this we first define a superpotential
\begin{equation}\label{superpotential}
W\equiv \sqrt{\frac{1}{3}P_{ij}P^{ij}} = \sqrt{\frac{2}{3} P^r P^r}
\end{equation}
The first term in \eqref{potential} can obviously be written using the superpotential. Less obviously, the last term can also be expressed using $W$ \cite{Ceresole:2001wi}
\begin{equation}
2\mN_{iA} \mN^{iA} =  \frac{3}{4} g^{XY} K_X K_Y=      \frac{9}{2}g^{XY}\pa_X W\pa_Y W
\end{equation}
where we have used
\begin{equation}\label{supPotId}
\pa_X W = \frac{2}{3W} P^r D_X P^r = \frac{2}{3W} P^r \mR^r_{XY} K^Y
\end{equation}
and \eqref{Ridentity}.
We would like to emphasize that the analysis above is general and indepedent of the spacetime metric.
In particular, it is valid for both compact and non-compact spacetimes.

By decomposing the prepotentials into their norms and phases
\begin{equation}
P^r = \sqrt{\frac{3}{2}}W Q^r, \qquad Q^rQ^r=1
\end{equation}
the contribution from the vector multiplet scalars can be brought to a similar form
\begin{equation}
2 P^r_x P^r_y g^{xy} = \frac{9}{2}g^{xy}\pa_x W\pa_y W+ \frac{9}{2} W^2 \left( \pa_x Q^r \right) \left( \pa^x Q^r \right)
\end{equation}
The potential is therefore given by \cite{LopesCardoso:2001rt,LopesCardoso:2002ec}
\begin{equation}
V= - 6W^2 +\frac{9}{2} g^{\Lambda\Sigma}\pa_{\Lambda}W\pa_{\Sigma}W + \frac{9}{2} W^2 \left( \pa_x Q^r \right) \left( \pa^x Q^r \right)
\end{equation}
where $g^{\Lambda\Sigma}$ is the metric of the complete scalar manifold.

When the phases $Q^r$ depend only on the quaternions
\begin{equation}\label{constraint}
\pa_x Q^r =0
\end{equation}
the potential takes the Bogomolnyi form
\begin{equation}\label{bogo}
V= - 6W^2 +\frac{9}{2} g^{\Lambda\Sigma}\pa_{\Lambda}W\pa_{\Sigma}W 
\end{equation}
An important implication of this analysis is that an $\mN=2$ supergravity theory without vector multiplets is described by the Bogomolnyi potential \eqref{bogo} and the superpotential \eqref{superpotential}. In particular, the theory that we study in section \ref{G2section} contains two hypermultiplets and no vector multiplets and therefore has a description in terms of an $\mN=1$ superpotential.
On the other hand, in section \ref{universalHyperSection} we study the theory with $n_V=n_H=1$, which, in general, does not admit the constraint \eqref{constraint}, and therefore its scalar potential cannot be brought to the Bogomolnyi form.
We find that only particular truncations of the theory, which correspond to flat-sliced domain walls, satisfy the condition \eqref{constraint}, in which case the potential can be written in the form \eqref{bogo}, but otherwise it is impossible.

\subsection{Domain walls with $S^4$ boundary}

The main purpose of this paper is to study domain wall solutions with an $S^4$ boundary metric. The five dimensional bulk metric is therefore given by
\begin{equation}
ds^2 = dr^2 +e^{2A(r)} ds^2_{S^4}
\end{equation}
where $ds^2_{S^4}$ is the metric of a unit four-sphere. The Ricci scalar and metric determinant are given by
\begin{equation}
\begin{aligned}
R &= e^{-2A(r)} R_{S^4} -20 A' \, ^2 -8 A'' \\
\sqrt{g} &= \sqrt{g_{S^4}} e^{4A(r)}
\end{aligned}
\end{equation}
where $R_{S^4} =48$ is the Ricci scalar of a unit four-sphere.

We now wish to study the equations of motion for configurations that preserve Euclidean symmetry on the four-sphere. Euclidean symmetry implies that the vector fields are set to zero and the scalars are functions of the radial coordinate only. The equations of motion that follow from the Lagrangian \eqref{Lagrangian} are then given by
\begin{equation}
\begin{aligned}\label{EOMs}
3 A'' + 6 A' \,^2 -3 e^{-2A} - \frac{1}{2} g_{\Lambda \Sigma} \pa_r \Phi^{\Lambda} \pa_r \Phi^{\Sigma} -V &= 0 \\
\pa_r \left( e^{4A} g_{\Lambda \Sigma} \pa_r \Phi^{\Sigma}   \right) + e^{4A} \pa_{\Lambda} V &=0
\end{aligned}
\end{equation}
where $\Phi^{\Lambda}$ stands for all the scalar fields.
In addition, the Einstein equations also imply the following constraint equation
\begin{equation}
6( A' \,^2 - e^{-2A})  + \frac{1}{2} g_{\Lambda \Sigma} \pa_r \Phi^{\Lambda} \pa_r \Phi^{\Sigma} - V =0
\end{equation}
For examples of domain-wall solutions in $\mN=2$ supergravity we refer to \cite{Behrndt:2002ee,Behrndt:2000km,Lukas:1998tt,Gunaydin:1999zx}.

\section{The Leigh-Strassler flow}\label{universalHyperSection}

As a warmup exercise, in this section we describe the Leigh-Strassler flow, where only one of the three chiral multiplets of $\mN=4$ SYM becomes massive.
The gravity dual of this theory is $\mN=2$ supergravity coupled to one vector multiplet and one hypermultiplet ($n_V=n_H=1$) \cite{Pilch:2000fu}.
We describe the universal hypermultiplet, which is part of this theory, in order to prepare the ground for the study of the theory coupled to two hypermultiplets in section \ref{G2section}.

The scalar manifold of the theory with one vector multiplet and one hypermultiplet is given by \cite{Pilch:2000fu}
\begin{equation}
\mM = O(1,1)\times \frac{SU(2,1)}{SU(2)\times U(1)}
\end{equation}
The first factor in $\mM$ is a "very special manifold" describing the one scalar in the vector multiplet
\begin{equation}
\phi = \beta
\end{equation}
The coset factor in $\mM$ is a quaternionic K\"{a}hler manifold describing the four scalars in the hypermultiplet
\begin{equation}
q^X = \left( z_1,z_2,\zb_1,\zb_2 \right)
\end{equation}
We start by describing the manifold and its isometries.
Then we describe the gauging of an Abelian $U(1)\times U(1)$ subgroup of the quaternionic manifold.

\subsection{The very special manifold}
The metric on the very special manifold is given by
\begin{equation}
g_{\beta\beta}=12
\end{equation}
The constants $C_{IJK}$ can be chosen to be all but $C_{011}$ equal to zero.
We can further impose that $a_{IJ}$ is diagonal, and together with the constraint to the surface \eqref{vectorSurface} we then get
\begin{equation}
\begin{aligned}
C_{011}&=\frac{\sqrt{3}}{2}, \qquad h^0 &= \frac{1}{\sqrt{3}} e^{-4\beta},\qquad  h^1 &= \sqrt{\frac{2}{3}} e^{2\beta} \\
a_{00} &= e^{8 \beta}, \qquad  a_{11}&=e^{-4\beta}, \qquad a_{01} &= 0
\end{aligned}
\end{equation}

\subsection{The universal hypermultiplet}

The K\"{a}hler potential on the quaternionic manifold is given by
\begin{equation}
\mK = - \log ( 1- |z_1|^2- |z_2|^2 )
\end{equation}
The kinetic term in our notations is than
\begin{equation}
\mL_{\text{kinetic}}=\mK_{\alpha \bar{\beta}} \mD_{\mu}z^{\alpha}\mD^{\mu}\zb^{\bar{\beta}}
= \frac{1}{2} g_{XY}\mD_{\mu}q^X \mD^{\mu}q^Y
\end{equation}
where
\begin{equation}
\mK_{\alpha \bar{\beta}} = \pa_{\alpha}\pa_{\bar{\beta}}  \mK
\end{equation}
The resulting manifold is the complex projective plane ($\mathbb{C}\mathbb{P}_2$). The Bergman metric on $\mathbb{C}\mathbb{P}_2$ is \cite{BrittoPacumio:1999sn,Behrndt:2000ph}
\begin{equation}
\mK_{\alpha \bar{\beta}}dz^{\alpha}d \zb ^{\bar{\beta}} = 
\frac{ \left| dz_1\right|^2  + \left|  dz_2 \right|^2 }{ 1- |z_1|^2- |z_2|^2}
+\frac{\left| \zb_1 dz_1+\zb_2 dz_2  \right|^2}{\left( 1- |z_1|^2- |z_2|^2\right)^2}
\end{equation}
We now introduce a polar system of coordinates on $\mathbb{C}\mathbb{P}_2$
\begin{equation}\label{projectivecoordinates}
\begin{aligned}
z_1 &=R \cos \left( \frac{\theta}{2}   \right)  e^{i\frac{\phi+\psi}{2}} \\
z_2 &= R \sin \left( \frac{\theta}{2}   \right)  e^{i\frac{-\phi+\psi}{2}}
\end{aligned}
\end{equation}
where all the field $q^{X}= \left(R,\theta,\phi,\psi\right)$ are real. It is most convenient to describe the quaternionic K\"{a}hler manifold in this system of coordinates. The K\"{a}hler metric is than given by
\begin{equation}
\mK_{\alpha \bar{\beta}}dz^{\alpha}d \zb ^{\bar{\beta}} = 
\frac{dR^2}{(1-R^2)^2}
+\frac{R^2}{4(1-R^2)}\left( \sigma_1^2+ \sigma_2^2  \right)
+\frac{R^2}{4(1-R^2)^2} \sigma_3^2
\end{equation}
where the $SU(2)$ one-forms are
\begin{equation}
\begin{aligned}
\sigma_1 &= + \cos\psi d \theta+\sin\psi\sin\theta d\phi \\
\sigma_2 &= -\sin\psi d \theta+\cos\psi\sin\theta d \phi \\
\sigma_3 &= d\psi +\cos \theta d\phi
\end{aligned}
\end{equation}
The metric can be written using the vielbeins
\begin{equation}
\begin{aligned}
\mK_{\alpha \bar{\beta}}dz^{\alpha}d \zb ^{\bar{\beta}}  =  \eta_{ab}e^a e^b = e^R e^R +e^1 e^1 +e^2 e^2 +e^3 e^3 \\
e^R = \frac{dR}{1-R^2}  ,\qquad
e^3 = \frac{R}{2(1-R^2)} \sigma_3 ,\qquad
e^{1/2} &= \frac{R}{2\sqrt{1-R^2}} \sigma_{1/2}
\end{aligned}
\end{equation}
One can then define the complex vielbeins
\begin{equation}
\begin{aligned}
u &= -\frac{R}{2\sqrt{1-R^2}} \left( \sigma_2+i\sigma_1 \right)  \\
v &=  +\frac{1}{1-R^2} \left( dR - i \frac{R}{2}\sigma_3  \right)
\end{aligned}
\end{equation}
and with $f^{iA} = f^{iA}_X dq^X$
\begin{equation}\label{comviel}
f^{iA}=
\left(
\begin{tabular}{cc}
$u$ & $-v$ \\
$\bar{v}$ & $\bar{u}$ 
\end{tabular}
\right), \qquad
f_{iA}=
\left(
\begin{tabular}{cc}
$\bar{u}$ & $-\bar{v}$ \\
$v$ & $u$ 
\end{tabular}
\right)
\end{equation}
in terms of which the metric is given by $g=f^{iA}\otimes f_{iA}=2 (u\bar{u}+v\bar{v})$.

Using the vielbeins we can derive the $SU(2)$ curvature
\begin{equation}
\mathcal{R}_i \,\! ^j = -\frac{1}{2}f_{iA}\wedge f^{jA} = -\frac{1}{2}
\left(
\begin{tabular}{cc}
$-(u\wedge\bar{u}+v\wedge\bar{v})$ & $2\bar{u}\wedge\bar{v}$ \\
$- 2u \wedge v$ & $(u\wedge\bar{u}+v\wedge\bar{v})$ 
\end{tabular}
\right)
\end{equation}
which can be decomposed into $SU(2)$ triplets \footnote{We follow the conventions of \cite{Ceresole:2001wi}. In order to translate to the conventions of \cite{Behrndt:2000ph}, one has to multiply the $SU(2)$ curvature $\mathcal{R}^i$ by 2, and the connections $\omega^i$ by -2.}
\begin{equation}
\begin{aligned}
\mathcal{R}^1 &=  \frac{R }{4 \left(1-R^2\right)^{3/2}} \left[2 \,  dR \wedge \sigma_1+R\; \sigma_2\wedge \sigma_3 \right]\\
\mathcal{R}^2 &= \frac{R}{4 \left(1-R^2\right)^{3/2}}  \left[-  2\, dR \wedge  \sigma_2+ R \; \sigma_1\wedge \sigma_3 \right] \\
\mathcal{R}^3 &= \frac{R }{4 \left(1-R^2\right)^2} \left[ 2\,  dR \wedge \sigma_3  +R \left(1-R^2\right) \sigma_1 \wedge \sigma_2 \right]
\end{aligned}
\end{equation}
The curvature triplets can be derived from the following $SU(2)$ connections (using equation \eqref{curvature})
\begin{equation}
\begin{aligned}
\omega^1 = + \frac{\sigma_1}{2\sqrt{1-R^2}}  , \qquad
\omega^2 = - \frac{\sigma_2}{2\sqrt{1-R^2}},\qquad
\omega^3 = +\frac{1}{4}  \frac{2-R^2}{1-R^2}   \sigma_3
\end{aligned}
\end{equation}

The isometry of this space is $SU(2,1)$.
The eight generators of $SU(2,1)$ can be classified as follows:
\begin{enumerate}
\item The generators of the compact subgroup $SU(2) \times U(1)$.
\item The generators of the non-compact coset $\frac{SU(2,1)}{SU(2)\times U(1)} $.
\end{enumerate}

Since eventually we will be interested in gauging compact isometries, we concentrate here on the generators of the compact subgroup $SU(2) \times U(1)$, which are given by the following Killing vectors
\begin{equation}\label{killing}
\begin{aligned}
SU(2) \quad  
&\begin{cases} 
k_1 &=   \sin \phi \,  \pa_{\theta}    + \cos \phi  \left(+  \cot \theta \,  \pa_{\phi}     -\csc \theta \, \pa_{\psi}    \right)\\
k_2 &=   \cos\phi  \, \pa_{\theta}   + \sin \phi  \left( -\cot\theta \,  \pa_{\phi}  +  \csc\theta \, \pa_{\psi}  \right)\\
k_3 &=  + \; \pa_{ \phi}
\end{cases}\\
U(1) \quad  
& \begin{cases} 
k_4 &= - \; \pa_{ \psi}
\end{cases}
\end{aligned}
\end{equation}
(for the more details about the $SU(2,1)$ algebra and its generators see appendix \ref{KillingApp}).
The action of the $SU(2)$ subgroup corresponds to "rotations" of the two complex coordinates $z_1,z_2$, and the three generators $\left( k_1,k_2,k_3 \right)$ fulfill the $SU(2)$ algebra $\left[k_m,k_n\right]=i\epsilon_{mnl}k_{l}$. $k_3$ is the generator of the Abelian subgroup inside $SU(2)$ which corresponds to translations in $\phi$. 
The $U(1)$ subgroup, represented by the generator $k_4$, corresponds to translations in $\psi$. These two Abelian $U(1)$ subgroups generate phase transformations in $z_1,z_2$, and are precisely the ones we want to gauge.

We end the discussion on the \textit{ungauged} quaternionic K\"{a}hler manifold with the prepotentials associated with the Killing vectors. The prepotentials can be derived using equation \eqref{prepotentialFormula}.
The prepotentials associated with the Killing vectors of the gauged isometries $k_3$ and $k_4$ are given by
\begin{equation}
p_3 = 
-\frac{1}{2\sqrt{1-R^2}}
\left(
\begin{tabular}{c}
$\sin\theta\sin\psi$ \\
$-\sin\theta\cos\psi$ \\
$\frac{2-R^2}{2\sqrt{1-R^2}}\cos\theta$
\end{tabular}
\right), \qquad
p_4 = 
\frac{R^2}{4(1-R^2)}
\left(
\begin{tabular}{c}
0 \\
0 \\
1
\end{tabular}
\right)
\end{equation}
The prepotentials associated with the rest of the isometries do not play any role here, but they can be derived in a similar way (and are given in appendix \ref{KillingApp} only for completeness).

\subsection{The gauging}

As explained at the end of the previous subsection, we want to gauge the Abelian $U(1)\times U(1)$ subgroup of $SU(4)\simeq SO(6)$
\begin{equation}
U(1)\times U(1) \subset  SU(2)\times U(1)\subset  SU(4)
\end{equation}
Along the flow both $U(1)$'s are in general broken\footnote{Particular solutions might still preserve the gauged isometries, or part of them, as will be discussed in section \ref{gaugeIsoSec}.}. In the UV fixed point both are preserved, since they are part of the $SU(4)$ symmetry of $\mN=4$ SYM. The Leigh-Strassler fixed point in the IR preserves only a linear combination $U(1)_V$ of them, while another combination $U(1)_B$ becomes massive and is therefore broken. $U(1)_V$ corresponds to the graviphoton $A^1$ and $U(1)_B$ corresponds to the gauge vector $A^0$.

From the field theory point of view, $U(1)_B$ corresponds to $U(1)_{56}$ and $U(1)_V$ correspond to the linear combination $U(1)_{12}+U(1)_{34}$, where $U(1)_{ij}$ represent $SO(2)$ rotations in the $i-j$ plane in $\mathbb{R}^6$. 

Next, we want to understand how the fields transform under the subgroups $U(1)_V$ and $U(1)_B$ of $SO(6)$. To do this, we group the $\mathbb{R}^6$ coordinates $x_1,\dots,x_6$ into three complex combinations
\begin{equation}
y_1 = x_1 + i x_2, \qquad
y_2 = x_3 + i x_4, \qquad
y_3 = x_5 + i x_6
\end{equation}
$z_1$ corresponds to the fermion mass term and therefore transform as the 3-form $d \bar{y}_1 \wedge d\bar{y}_2 \wedge d y_3$.
$z_2$ corresponds to the boson mass term and therefore transform as $y_3^2$. Therefore the charges of the fields under rotations in $\mathbb{R}^6$ are given by the values in table \ref{charges}.

\begin{table}[t]
\vspace{0 mm}
\begin{center}
\def\arraystretch{1.7}
\begin{tabular}{ |c|c|c|c| } 
 \hline
& $U(1)_{12}$ & $U(1)_{34}$ & $U(1)_{56}$ \\ 
  \hline
$z_1$ & $-1$ & $-1$ & $+1$ \\ 
  \hline
$z_2$ & $0$ & $0$ & $+2$ \\ 
 \hline
\end{tabular}
\caption{The charges of the fields under rotations in $\mathbb{R}^6$.}
\label{charges}
\end{center}
\end{table}%

The kinetic term in the bulk is therefore given by
\begin{equation}\label{gaugekin}
\begin{aligned}
\mL_{\text{kinetic}} &= \frac{1}{2} \left[ (d A^0 )^2  +2 (d A^1 )^2  \right] \\
&+
\left| \left( \pa-iA^0 +2i A^1  \right) z_1 \right|^2
+\left| \left(  \pa- 2i A^0 \right) z_2 \right|^2
+\dots
\end{aligned}
\end{equation}
The factor of $2$ in front of $(d A^1 )^2$ is due to the fact that the graviphoton $A^1$ corresponds to the diagonal combination $U(1)_{12}+U(1)_{34}$.
Normalizing $A^0$ as in \eqref{Lagrangian} (i.e. $A^0 \rightarrow \sqrt{2} A^0$ in \eqref{gaugekin}) and changing coordinates to \eqref{projectivecoordinates} we then have
\begin{equation}
\begin{aligned}
\mL_{\text{kinetic}} &=  (d A^0 )^2  +(d A^1 )^2 \\
&
+\frac{1}{2}\left| (\pa \phi)^2 - \sqrt{2}  i A^0 - 2i A^1  \right|^2\\
&+\frac{1}{2}\left| (\pa \psi)^2 + 3 i \sqrt{2} A^0 - 2i A^1  \right|^2
+\dots
\end{aligned}
\end{equation}
The fields $R$ and $\theta$ are not charged under the gauge groups.
The Killing vectors of the gauged isometries $K^X_I(q)$ are therefore given by
\begin{equation}\label{Killingvec}
\vec{K}_0 =
\sqrt{2}
 \left(
\begin{tabular}{c}
0 \\
0 \\
1 \\
-3
\end{tabular}
\right),
\qquad\qquad
\vec{K}_1 =
2 
 \left(
\begin{tabular}{c}
0 \\
0 \\
1 \\
1
\end{tabular}
\right)
\end{equation}
We can express this result in differential form and using the Killing isometries of the manifold \eqref{killing}
\begin{equation}
\begin{aligned}
K_0 &= K_0^X dq^X = \sqrt{2} \left( d\phi -3 d\psi \right) = \sqrt{2} \left(  k_3 +3 k_4 \right) \\
K_1 &= K_1^X dq^X = 2 \left( d\phi + d\psi \right) = 2 \left(  k_3 - k_4 \right)
\end{aligned}
\end{equation}
The corresponding prepotentials $P_I^r$ are then given by the same combinations of the associated prepotentials $p_3$ and $p_4$
\begin{equation}\label{prepotentials}
\begin{aligned}
P_0 ^r&=   \sqrt{2} \left(  p_3 +3 p_4 \right)= 
\sqrt{2} \left(
\begin{tabular}{c}
$\frac{\sin \theta  \sin \psi }{ \sqrt{1-R^2}}, \quad
-\frac{\sin \theta  \cos \psi }{ \sqrt{1- R^2}}, \quad
\frac{   \left(2-R^2\right) \cos \theta  -3 R^2}{2 \left(1- R^2\right)}$
\end{tabular}
\right)
\\
P_1^r  &=   2 \left(  p_3 - p_4 \right)=
2\left(
\begin{tabular}{c}
$\frac{\sin \theta \sin \psi }{ \sqrt{1-R^2}} , \quad 
-\frac{\sin \theta  \cos \psi }{\sqrt{1-R^2}}, \quad
\frac{\left(2-R^2\right) \cos \theta +R^2}{2(1- R^2)}$
\end{tabular}
\right)
\end{aligned}
\end{equation}
Now we basically have all the information needed to evaluate the potential  \eqref{potential} and BPS equations \eqref{susyVar}, but before doing so we first want to discuss some aspects of the gauging.

\subsection{A different system of coordinates}

At this point we would like to make a connection with another system of coordinates that appear in the literature
\begin{equation}
\begin{aligned}
z_1 &= -i \frac{z-\zb}{1+|z|^2}e^{i\frac{\phi+\psi}{2}} \\
z_2 &= \frac{z+\zb}{1+|z|^2}e^{i\frac{-\phi+\psi}{2}}
\end{aligned}
\end{equation}
This system of coordinates is very similar to the polar system of coordinates \eqref{projectivecoordinates} - $\phi$ and $\psi$ are defined in the same way, while $R$ and $\theta$ are related to $z$ and $\zb$ by
\begin{equation}\label{changeofvariables}
R= \frac{2 |z|}{1+|z|^2}, \qquad \tan\frac{\theta}{2} = i \frac{z+\zb}{z-\zb}
\end{equation}

\subsection{The gauged isometries}\label{gaugeIsoSec}

The Abelian gauge group $U(1)\times U(1)$ is completely broken along the flow.
This can be understood by examining the mechanism that gives mass to the vector fields.
A vector mass term can come from the kinetic term of the hypermultiplet scalars, which takes the form \eqref{Lagrangian}
\begin{equation}
-\frac{1}{2}\left( \pa_{\mu} q^X + g^2 A_{\mu}^I K_I^X  \right)^2
\end{equation}
We see that, due to the gauge covariant derivative, a vector mass term is generated $g^2 A_{\mu}A^{\mu} \left| K \right|^2$, where $A_{\mu}$ is in general a linear combination of the gauge fields and $K^X$ is the corresponding Killing vector.
The vector mass is then proportional to
\begin{equation}
m^2_A \sim \left|K\right|^2
\end{equation}
where $\left|K\right|$ is the norm of the corresponding Killing vector $\left|K\right|^2 \equiv g_{XY}K^XK^Y$.
Whenever $\left|K\right|\neq0$, the corresponding vector field is massive and as a consequence the gauge group associated with it is broken.
Whenever $\left|K\right|=0$, on the other hand, the corresponding vector field remains massless and the associated gauge group is preserved.

To understand how the gauge group is broken we therefore have to evaluate the norm of the Killing vectors in our theory
\begin{equation}\label{Knorms}
\begin{aligned}
\left| K_0\right| ^2 &= R^2\frac{ 10  - 6 \cos\theta -R^2 \sin ^2 \theta }{\left(1-R^2\right)^2}
=\frac{4 \left(3 z \zb^2+z+\zb^3+3 \zb\right) \left(z \left(z^2+3 z \zb+3\right)+\zb\right)}{(1-|z|^2)^4}
\\
\left| K_1\right| ^2 &=2 R^2 \frac{2+2 \cos\theta-R^2 \sin ^2\theta }{\left(1-R^2\right)^2}
= -\frac{8 \left(1-z^2\right) \left(1-\zb^2\right) }{L^2 (1-|z|^2)^4}(z-\zb)^2
\\
\left| K_R\right| ^2 &=8 R^2\frac{  2-2 \cos \theta -R^2 \sin ^2\theta}{\left(1-R^2\right)^2}
 =\frac{32 \left(1+z^2\right) \left(1+\zb^2\right) }{L^2 (1-|z|^2)^4}(z+\zb)^2
 \end{aligned}
\end{equation}
where we have defined
\begin{equation}
K_R \equiv \sqrt{2} K_0+K_1
\end{equation}
for reasons that will become clear shortly.
Along the flow, both $\left| K_0\right|$ and $\left| K_1\right|$ are non-zero, and therefore $U(1)\times U(1)$ is completely broken.

Let us now examine the behavior at the fixed points.
The UV and Leigh-Strassler IR fixed points are located at
\begin{equation}
\begin{aligned}
\text{UV fixed point:}& 
\qquad
z =\zb = 0
&\Longleftrightarrow	&
\qquad
R=0
\\
\text{LS IR fixed point:}& 
\qquad 
z =-\zb =i \sqrt{7+4 \sqrt{3}}
&\Longleftrightarrow	&
\qquad
\begin{cases}
R &= \frac{\sqrt{7+4 \sqrt{3}}}{4+2 \sqrt{3}} \qquad\\
 \theta &= 0
\end{cases}
\end{aligned}
\end{equation}
The values of the norms of $K_0$,$K_1$ and $K_R$ at these points are presented in table \ref{gaugedIso}.
At the UV fixed point both  $K_0$ and $K_1$ are massless, as expected, since the corresponding $U(1)\times U(1)$ gauge group is part of the $SU(4)\simeq SO(6)$ symmetry group of $\mathcal{N}=4$ SYM.
At the Leigh-Strassler fixed point in the IR, on the other hand, both of them become massive. However, the linear combination $K_R$ remains massless.
This means that the Leigh-Strassler fixed point preserves a residual $U(1)_R\subset SU(4)$ symmetry corresponding to the linear combination $K_R$, as expected.

\begin{table}[t]
\vspace{0 mm}
\begin{center}
\def\arraystretch{1.7}
\begin{tabular}{ |c|c|c| } 
 \hline
& \; UV \; & \; IR \;  \\ 
  \hline
$\left| K_0\right| ^2$ & $0$ & $\frac{16}{9}$  \\ 
  \hline
$\left| K_1\right| ^2$ & $0$ & $\frac{32}{9}$ \\ 
  \hline
$\left| K_R\right| ^2$ &$0$ &$0$ \\
 \hline
\end{tabular}
\qquad\qquad
\def\arraystretch{1.9}
\begin{tabular}{|c|c|c| } 
 \hline
 & $\theta=0$ \; or \; $\zb=-z$&  $\theta=\pi$ \; or \;  $\zb=z$ \\ 
  \hline
$\left| K_0\right| ^2$  & $\frac{4 R^2}{\left(1-R^2\right)^2}$ 
&$\frac{16 R^2}{\left(1-R^2\right)^2}$ \\  [1ex]
  \hline
$\left| K_1\right| ^2$ & $ \frac{8 R^2}{\left(1-R^2\right)^2}$ & $0$ \\  [1ex]
  \hline
$\left| K_R\right| ^2$  & $0$ & $\frac{32 R^2}{\left(1-R^2\right)^2}$ \\ [1ex]
 \hline
\end{tabular}
\caption{Left: The values of the norms of the Killing vectors at the UV and IR fixed points. While at the UV fixed point both $U(1)$'s are preserved, at the IR fixed point only the linear combination $U(1)_R$ is preserved.
Right: The norms of the Killing vectors in two flat-sliced domain wall truncations.
In the first truncation $\theta=0$ the isometry $U(1)_R$ is preserved all along the flow, while in the second truncation $U(1)_V$ is preserved along the flow.}
\label{gaugedIso}
\end{center}
\end{table}%

Finally, we would like to consider two truncations to flat-sliced domain walls, as suggested by the result \eqref{Knorms}.
The first truncation is set by $\zb=-z$ (or $\theta=0$), and corresponds to the FGPW flow in flat spacetime.
It is evident that while $K_0$ and $K_1$ become massive, the diagonal combination $K_R$ remains massless all along the flow.
As expected, the FGPW flow therefore preserves a residual $U(1)_R\subset SU(4)$ symmetry.
The second truncation is set by $\zb=z$ (or $\theta=\pi$).
This flow preserves the $U(1)_V$ part of $SU(4)$ which corresponds to $K_1$.
Note that $U(1)_B$, which is associated with $K_0$, is always broken (except for the UV fixed point), and hence deserves the subscript $B$ $(\ddot\smile)$.

\subsection{The scalar potential}

Finally, we have all the information needed to evaluate the scalar potential \eqref{potential}.
Using the complex vielbeins \eqref{comviel} and the Killing vectors of the gauged isometries \eqref{Killingvec} we can evaluate $\mN^{iA}$
\begin{equation}
\mN^{iA}=
- e^{-4\beta} \frac{R}{4\sqrt{1-R^2}} \left(
\begin{array}{cc}
 e^{i \psi  }\left(1+2 e^{6 \beta }\right)  \sin \theta
 & i  \frac{ 2 e^{6 \beta } (\cos \theta +1)+\cos \theta  -3  }{ \sqrt{1-R^2}} \\
i  \frac{ 2 e^{6 \beta } (\cos \theta +1)+\cos \theta  -3  }{ \sqrt{1-R^2}} 
 &  e^{i \psi  }\left(1+2 e^{6 \beta }\right)  \sin \theta\\
\end{array}
\right)
\end{equation}
The last term in the potential \eqref{potential} is therefore given by
\begin{equation}
\mN^{iA} \mN_{iA}
=
\frac{e^{-8 \beta } R^2}{8 (1-  R^2)^2}
\left[ \left(2 e^{6 \beta } (\cos \theta+1)+\cos \theta -3\right)^2
+(1+ 2 e^{6 \beta })^2 \left(1-R^2\right) \sin ^2\theta  \right]
\end{equation}
The first two terms in \eqref{potential} are simple functions of the prepotentials we found \eqref{prepotentials}. Plugging it all together we find
\begin{equation}
\begin{aligned}
V &=
R^2 
\frac{10 -  6 \cos \theta  -  R ^2 \sin ^2\theta    }{4 \left(1-R^2\right)^2}
e^{-8 \beta }
-2 \frac{ 2+ R^2 (1-\cos \theta )}{1-R^2}e^{-2 \beta } 
-\frac{ 2(1-2R^2)+   \sin ^2 \frac{\theta }{2} (\cos \theta +3)R^4   }{ \left(1-R^2\right)^2}e^{4 \beta } \\
\end{aligned}
\end{equation}
Changing variables to \eqref{changeofvariables}
\begin{equation}
\begin{aligned}
V &=
\frac{|z_1|^2+4 |z_2|^2 - |z_1|^2|z_2|^2}{(1- |z_1|^2-|z_2|^2)^2}
e^{-8\beta}
-\frac{4(1+|z_2|^2)}{1- |z_1|^2-|z_2|^2}
e^{-2\beta}
+2\left(
-1+\frac{|z_1|^4}{(1- |z_1|^2-|z_2|^2)^2}
\right)
e^{4\beta}
\end{aligned}
\end{equation}
The equations of motion that result from this potential imply that both phases $\phi,\psi$ are constants.
The case with constant phases was studied in \cite{Bobev:2016nua}.

\section{The $G_{2,2}/SU(2)\times SU(2)$ coset model}\label{G2section}

We now turn to study the main objective of this paper, which is the gravity dual of the $\mN=1^*$ theory with masses
\begin{equation}
\begin{aligned}
m_1 =m_2 &=m_3 = m \\
\mt_1 = \mt_2 &= \mt_3 =\mt \\
m &\neq \mt
\end{aligned}
\end{equation}
In this case the theory is invariant under a global $SO(3)$ symmetry group.

In general, the superpotential of the $\mN=1^*$ theory is given by the following expression
\begin{equation}\label{superP}
\mW = \sqrt{2} g_{YM} f^{abc} Z_1^a Z_2^b Z_3^c +\frac{1}{2} \sum_{i=1}^3 m_{ij} Z_i ^a Z_j^a
\end{equation}
The first term in \eqref{superP} preserves the full $SU(4)$ R-symmetry of $\mN=4$ SYM, although only the subset $SU(3)\times U(1)_R \subset SU(4)$ is manifest. The mass term breaks, in general, the $SU(3)$ symmetry, leaving only a $U(1)_R$ factor inside $SU(4)$ unbroken. However, in the case $m_{ij} = m \delta_{ij}$, the subgroup $SO(3) \subset SU(3)$ is also preserved.
We are therefore interested in the decomposition
\begin{equation}
SU(4) \simeq SO(6) \rightarrow SO(3)\times U(1)_R
\end{equation}
$SO(3)$ can also be thought of as the diagonal subgroup of $SO(3)\times SO(3) \subset SO(6)$
\begin{equation}
\left(
\begin{array}{cc}
\multirow{ 2}{*}{$\quad SO(6)$} \\
&
\end{array}
\right)
=
\left(
\begin{array}{c|c}
$SO(3)$ & $0$ \\
\hline
$0$ & $SO(3)$
\end{array}
\right)
\end{equation}

The spectrum of deformations of the theory is classified by their transformation properties under the global symmetries.
To understand this classification, let us recall how $SU(4)$ representations decompose under $SU(4)\rightarrow SO(3)\times U(1)_R$:
\begin{equation}
\begin{aligned}
\b[4] & \rightarrow	\bs[3,1] +\bs[1,-3]		\\
\b[6] & \rightarrow	\bs[3,2] + \bs[3,-2]		\\
\b[10] & \rightarrow  \bs[5,2] +\bs[3,-2] + \bs[1,2] +\bs[1,-6]	\\
\b[20]' &\rightarrow  \bs[5,4] + \bs[5,-4] +\bs[5,0]+\bs[3,0]+\bs[1,4]+\bs[1,-4]	
\end{aligned}
\end{equation}
The notation for the decomposition is ${\bold J}_{Q}$, where ${\bold J}$ is the $SO(3)$ representation and $Q$ is related to the $U(1)_R$ charge by $R= -\frac{1}{3} Q$.
The $\b[4]$ and $\b[10]$ are complex and therefore the spectrum also contains their complex conjugates $\bb[4]$ and $\bb[10]$. The $\b[6]$ and the $\b[20]'$ are real.

Now we can make the connection with the spectrum of operators that was discussed in section \ref{FieldTheory}.
The $\bs[1,\mp 6]$ inside $\b[10]$ and $\bb[10]$ are the left-handed and right-handed gaugino bilinears.
The $\bs[1,\pm 2]$ inside $\b[10]$ and $\bb[10]$ are the fermion blinears.
The $\bs[1,\pm 4]$ inside $\b[20]'$ are the scalar deformations.
Together with the gauge kinetic term and the $\theta$-term, which are dual to the dilaton and the axion, and are singlets under $SU(4)$, we have eight scalar deformations.

The gravity dual of this theory is the $SO(3)$-invariant sector of $\mN=8$ supergravity, which can be consistently truncated to $\mN=2$ gauged supergravity coupled to $n_H=2$ hypermultiplets and no vector multiplets $n_V=0$ \cite{Pilch:2000fu}.
The scalar fields parameterize a quaternionic manifold given by the coset
\begin{equation}
\mM=\mQ = \frac{G_{2,2}}{SU(2)\times SU(2)}   \subset
\frac{E_{6,6}}{USp(8)} = \mM_{\mN=8}
\end{equation}
where $G_{2,2}$ is the non-compact form of the exceptional Lie group $G_2$.
See \cite{Pilch:2000fu,Ferrara:1989ik,Bodner:1989cg,Pilch:2000ue,Bianchi:2000sm,Suh:2011xc} for more works in the subject.

\subsection{Group theory}

In this subsection we describe the group $G_{2,2}$, following \cite{Gunaydin:2007qq} (see also \cite{Berkooz:2008rj,Chiodaroli:2008rj,Chiodaroli:2009cz}).
The fourteen generators of $G_{2,2}$ are given by
\begin{equation}
  \left(  E,H,F,Y_0,Y_{\pm}  ,E_{q_I},E_{p^I},F_{q_I},F_{p^I} \right) ,\qquad I=0,1
\end{equation}
They obey the universal commutation relations
\begin{equation}
\begin{aligned}
& [E,F]=H,           \qquad    &&     [H,E]=2E,       \qquad   &&   [H,F]=-2F,   \\
& [E_{p^I},E_{q_J}] = -2 \delta^I_J E,   \qquad  &&   [F_{p^I},F_{q_J}] = 2 \delta^I_J F ,  \\
&[E_{p^I},E_{p^J}] = 0,   \qquad &&   [F_{p^I},F_{p^J}] = 0,   \qquad   &&  [E_{q_I},E_{q_J}] = 0,   \qquad  &&   [F_{q_I},F_{q_J}] = 0,  \\
&[H,E_{p^I}] = E_{p^I},   \qquad &&   [H,F_{p^I}] = -F_{p^I},   \qquad  &&  [H,E_{q_I}] = E_{q_I},   \qquad   &&  [H,F_{q_I}] = -F_{q_I}, \\
&[F,E_{p^I}] = -F_{q_I},   \qquad &&   [E,F_{q_I}] = -F_{p^I},   \qquad  &&  [F,E_{q_I}] = F_{p^I},   \qquad   &&  [E,F_{p^I}] = E_{q_I}
\end{aligned}
\end{equation}
the $SL(2,\mathbb{R})$ sub-algebra
\begin{equation}
[Y_0,Y_{\pm}] = \pm Y_{\pm},   \qquad
[Y_-,Y_+] = Y_0
\end{equation}
under which $E$ and $F$ are singlets
\begin{equation}
[Y_0,E] = [Y_{\pm},E] = [Y_0,F] = [Y_{\pm},F] = 0
\end{equation}
and $E_{p,q}$ and $F_{p,q}$ transform as a spin $3/2$
\begin{equation}
\begin{aligned}
&\left[
Y_0,
\left(
\begin{tabular}{c}
$E_{p^0}$ \\
$E_{p^1}$ \\
$E_{q_1}$ \\
$E_{q_0}$
\end{tabular}
\right)
\right]
=\frac{1}{2}
\left(
\begin{tabular}{c}
$3E_{p^0}$ \\
$E_{p^1}$ \\
$-E_{q_1}$ \\
$-3E_{q_0}$
\end{tabular}
\right),
\qquad
&&\left[
Y_0,
\left(
\begin{tabular}{c}
$F_{p^0}$ \\
$F_{p^1}$ \\
$F_{q_1}$ \\
$F_{q_0}$
\end{tabular}
\right)
\right]
=\frac{1}{2}
\left(
\begin{tabular}{c}
$-3F_{p^0}$ \\
$-F_{p^1}$ \\
$F_{q_1}$ \\
$3F_{q_0}$
\end{tabular}
\right),
\\
&\left[
Y_+,
\left(
\begin{tabular}{c}
$E_{p^0}$ \\
$E_{p^1}$ \\
$E_{q_1}$ \\
$E_{q_0}$
\end{tabular}
\right)
\right]
=
\left(
\begin{tabular}{c}
$0$ \\
$\sqrt{\frac{3}{2}}E_{p^0}$ \\
$-\sqrt{2}E_{p^1}$ \\
$-\sqrt{\frac{3}{2}}E_{q_1}$
\end{tabular}
\right),
\qquad
&&\left[
Y_+,
\left(
\begin{tabular}{c}
$F_{p^0}$ \\
$F_{p^1}$ \\
$F_{q_1}$ \\
$F_{q_0}$
\end{tabular}
\right)
\right]
=
\left(
\begin{tabular}{c}
$-\sqrt{\frac{3}{2}}F_{p^1}$ \\
$\sqrt{2} F_{q_1}$ \\
$\sqrt{\frac{3}{2}}F_{q_0}$ \\
$0$
\end{tabular}
\right),
\\
&\left[
Y_-,
\left(
\begin{tabular}{c}
$E_{p^0}$ \\
$E_{p^1}$ \\
$E_{q_1}$ \\
$E_{q_0}$
\end{tabular}
\right)
\right]
=
\left(
\begin{tabular}{c}
$-\sqrt{\frac{3}{2}}E_{p^1}$ \\
$\sqrt{2}E_{q_1}$ \\
$\sqrt{\frac{3}{2}}E_{q_0}$ \\
$0$
\end{tabular}
\right),
\qquad
&&\left[
Y_-,
\left(
\begin{tabular}{c}
$F_{p^0}$ \\
$F_{p^1}$ \\
$F_{q_1}$ \\
$F_{q_0}$
\end{tabular}
\right)
\right]
=
\left(
\begin{tabular}{c}
$0$ \\
$\sqrt{\frac{3}{2}}F_{p^0}$ \\
$-\sqrt{2} F_{p^1}$ \\
$-\sqrt{\frac{3}{2}}F_{q_1}$
\end{tabular}
\right)
\end{aligned}
\end{equation}
Finally, they also obey the following commutation relations
\begin{equation}
\begin{aligned}
[E_{p^0},F_{p^0}] & = H+2Y_0,  	  	      \qquad    &&  [E_{q_0},F_{q_0}]  = H-2Y_0 ,\\
[E_{p^1},F_{p^1}] &= \frac{1}{3}(3H+2Y_0)   ,  \qquad    &&  [E_{q_1},F_{q_1}] = \frac{1}{3}(3H-2Y_0)   ,  \\
[E_{p^1},F_{q_1}] &= -\frac{4\sqrt{2}}{3} Y_+ , \qquad    &&  [E_{q_1},F_{p^1}] = +\frac{4\sqrt{2}}{3} Y_-
\end{aligned}
\end{equation}

\subsection{The maximal (compact) subgroup $SU(2)\times SU(2)$}
\begin{figure}
\begin{center}
\includegraphics[scale=0.7]{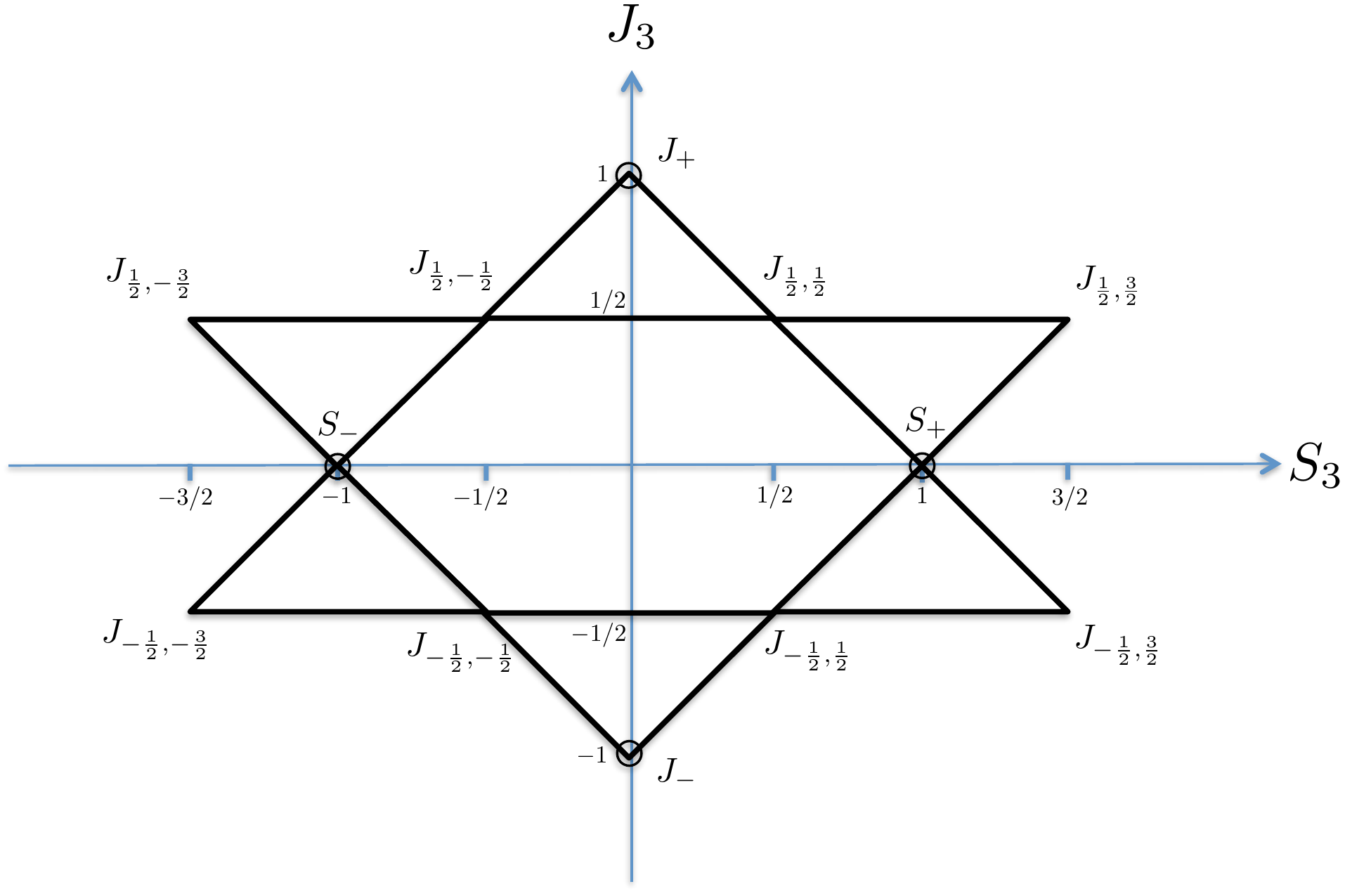}
\caption{Root diagram of the group $G_{2,2}$. The compact roots are indicated with a circle. The subscript on $J$ denotes the eigenvalues under $(-iJ_3,-iS_3)$.}
\label{RootDiagramG2}
\end{center}
\end{figure}

The maximal subgroup of the coset is the compact group $SU(2)\times SU(2)$.
Let us introduce a basis that manifest the compact generators:
\begin{equation}\label{J-generators}
\begin{aligned}
J_1  &=   \frac{1}{4 \sqrt{2}} \left(   - E_{p^0}-\sqrt{3} E_{q_1} + F_{p^0}  + \sqrt{3}  F_{q_1}      \right) \\
J_2  &=   \frac{1}{4 \sqrt{2}} \left(  -\sqrt{3} E_{p^1}+ E_{q_0}+\sqrt{3} F_{p^1}-  F_{q_0}     \right) \\
J_3  &=   \frac{1}{4}(F-E)+\frac{1}{2 \sqrt{2} } ( Y_+  +  Y_-)
\end{aligned}
\end{equation}
\begin{equation}\label{S-generators}
\begin{aligned}
S_1  &=    \frac{1}{4} \sqrt{\frac{3}{2}} \left(-\sqrt{3}  E_{p^0}+E_{q_1}+\sqrt{3} F_{p^0}- F_{q_1}\right)     \\
S_2  &=    \frac{1}{4} \sqrt{\frac{3}{2}} \left( E_{p^1}+\sqrt{3} E_{q_0}- F_{p^1}-\sqrt{3} F_{q_0}\right)    \\
S_3  &=   \frac{3}{4}(F-E)-\frac{1}{2 \sqrt{2} } ( Y_+  +  Y_-)
\end{aligned}
\end{equation}
Both sets of generators, $J_i$ and $S_i$ with $i=1,2,3$, separately obey the $SU(2)$ algebra
\begin{equation}
\begin{aligned}
\left[ J_i,J_j \right] &= -\ep _{ijk} J_k  \\
\left[ S_i,S_j \right] &= -\ep _{ijk} S_k  
\end{aligned}
\end{equation}
The two sets of $SU(2)$'s commute with each other
\begin{equation}
[J_i,S_j]=0
\end{equation}
We also define
\begin{equation}
\begin{aligned}
J_{\pm} &= J_1 + i J_2  \\
S_{\pm} &= S_1 + i S_2
\end{aligned}
\end{equation}
The root diagram of the group $G_{2,2}$ is described in figure \ref{RootDiagramG2}.

\subsection{Non-compact generators}
In addition to the 6 compact generators of the coset, there are also 8 non-compact generators given by
\begin{equation}
\begin{aligned}
J_{\frac{1}{2},-\frac{3}{2}} &=   (J_{-\frac{1}{2},\frac{3}{2}})^*  =  \frac{-E_{p^0}-i \sqrt{3} (E_{p^1}-i E_{q_1})+i E_{q_0}- F_{p^0}-i \sqrt{3} ( F_{p^1}-i F_{q_1})-i F_{q_0}}{2\sqrt{2}} \\
J_{\frac{1}{2},-\frac{1}{2}} &=   (J_{-\frac{1}{2},\frac{1}{2}})^*  =  \sqrt{\frac{2}{3}} \left(  Y_p-  Y_m      -i \sqrt{2} \, Y_0   \right) \\
J_{\frac{1}{2},\frac{1}{2}} &=   (J_{-\frac{1}{2},-\frac{1}{2}})^*  =  \frac{\sqrt{3} ( E_{p^0}+i E_{q_0})+i (E_{p^1}+i E_{q_1})+\sqrt{3} (F_{p^0}+i F_{q_0})+i (F_{p^1}+i F_{q_1})}{2 \sqrt{2}} \\
J_{\frac{1}{2},\frac{3}{2}} & =  (J_{-\frac{1}{2},-\frac{3}{2}} )^*  =-E -F -i H
\end{aligned}
\end{equation}

\subsection{Parameterization of the coset model}

We now describe the eight dimensional coset model $G_{2,2}/SU(2)\times SU(2)$ using the coordinates $\left(\tau_1,\tau_2,\zeta^0,\zeta^1,\zetat_0,\zetat_1,U,\sigma\right)$. 
We define $\tau=\tau_1+i \tau_2$. The metric on the scalar manifold (corresponding to the metric $g_{XY}$ in the notation of \eqref{Lagrangian}) is then
\begin{equation}
ds^2 = 2(u \ubar  +v \vbar  +e^1 \ebar ^1  + E_1 \Ebar _1 )
\end{equation}
where the vielbeins are given by
\begin{equation}
\begin{aligned}
u &=  \frac{e^{-U}}{2\sqrt{2} \tau_2^{3/2}}	\left(    d \zetat_0 +\tau d \zetat_1   +3 \tau^2 d\zeta^1 - \tau^3 d\zeta^0  \right)	\\
v &=	 dU +\frac{i}{2} e^{-2U}  \left(  d\sigma -\zeta^0 d\zetat_0 -\zeta^1 d\zetat_1 +\zetat_0 d\zeta^0 +\zetat_1 d\zeta^1   \right)   	\\
e^1 &= \frac{i\sqrt{3}}{2\tau_2} d\tau	\\
E_1 &= 	-\frac{e^{-U}}{2\sqrt{6} \tau_2^{3/2}}    \left(  3d\zetat_0 +  (\taub+2\tau)d\zetat_1
+3\tau (2\taub+\tau) d\zeta^1 - 3\taub\tau^2 d\zeta^0  \right)
\end{aligned}
\end{equation}
The complexified vielbeins are therefore
\begin{equation}
f^{iA}=
\left(
\begin{tabular}{cccc}
$u$ & $-v$ &   $e^1$   &  $-E_1$   \\
$\bar{v}$ & $\bar{u}$     &    $\bar{E}_1$    &    $\bar{e}^1$ 
\end{tabular}
\right), \qquad
f_{iA}=
\left(
\begin{tabular}{cccc}
$\bar{u}$ & $-\bar{v}$   &     $\bar{e}^1$     &     $-\bar{E}_1$    \\
$v$ & $u$      &     $E_1$     &     $e^1$
\end{tabular}
\right)
\end{equation}
in terms of which the metric is given by $g=f^{iA}\otimes f_{iA}=2 (u\bar{u}+v\bar{v}+e^1 \ebar ^1  + E_1 \Ebar _1)$.

Using the vielbeins we can derive the $SU(2)$ curvature
\begin{equation}
\begin{aligned}
\mathcal{R}_i \,\! ^j &= -\frac{1}{2}f_{iA}\wedge f^{jA} = \\
&-\frac{1}{2}
\left(
\begin{tabular}{cc}
$-(u\wedge\bar{u}+v\wedge\bar{v} +E_1 \wedge \bar{E}_1 +e^1 \wedge \bar{e}^1    )$ & $2 (\bar{u}\wedge\bar{v}   + \bar{E}_1 \wedge \bar{e}^1 )$ \\
$- 2(u \wedge v +E_1 \wedge e^1 )$ & $(u\wedge\bar{u}+v\wedge\bar{v} +E_1 \wedge \bar{E}_1 +e^1 \wedge \bar{e}^1  )$ 
\end{tabular}
\right)
\end{aligned}
\end{equation}
which can be decomposed into $SU(2)$ triplets
\begin{equation}
\begin{aligned}
\mathcal{R}^1 &=  -\frac{i}{2}\left( u\wedge v  - \bar{u} \wedge \bar{v}  +E_1 \wedge e^1   -   \bar{E}_1  \wedge  \bar{e}^1   \right)		  \\
\mathcal{R}^2 &=  -\frac{1}{2}\left( u\wedge v  + \bar{u} \wedge \bar{v}  + E_1  \wedge  e^1  +  \bar{E}_1  \wedge    \bar{e}^1  \right)		 \\
\mathcal{R}^3 &=   -\frac{i}{2}\left( u\wedge \bar{u}   +v \wedge \bar{v}  + E_1 \wedge \bar{E}_1 +e^1 \wedge \bar{e}^1  \right)	
\end{aligned}
\end{equation}
The curvature triplets can be derived from the following $SU(2)$ connections (using equation \eqref{curvature})
\begin{equation}
\begin{aligned}
\omega^1 = -\frac{i}{2} (u- \bar{u}),			 \qquad
\omega^2 = -\frac{1}{2} (u + \bar{u}),			\qquad
\omega^3 =   -\frac{i}{4}(v-\bar{v}) + \frac{i\sqrt{3}}{4} (e^1 -\bar{e}^1)
\end{aligned}
\end{equation}

Using this parameterization, the Killing vectors of $G_{2,2}$ take the form
\begin{equation}
\begin{aligned}
E &= 
\left( 0,0,0,0,0,0,0,1 \right) , 
 &H&= 
\left(
0,0,-\zeta ^0,-\zeta ^1,-\tilde{\zeta }_0,-\tilde{\zeta }_1,-1,-2 \sigma
\right)
\\
E_{p^0} &=
\left(
0,0,0,0,1,0,0,-\zeta ^0
\right) ,
&E_{q_0}&=
\left(
0,0,-1,0,0,0,0,-\tilde{\zeta }_0
\right)
\\
E_{p^1} &=
\left(
0,0,0,0,0,\sqrt{3},0,-\sqrt{3} \zeta ^1
\right) ,
&E_{q_1} &=
-\frac{1}{\sqrt{3}}
\left(
0,0,0,1,0,0,0,\tilde{\zeta }_1
\right)
\end{aligned}
\end{equation}
\begin{equation}
\begin{aligned}
Y_0 &= 
\frac{1}{2}
\left(
-2 \tau _1,-2 \tau _2,3 \zeta ^0,\zeta ^1,-3 \tilde{\zeta }_0,-\tilde{\zeta }_1,0,0
\right)
\\
Y_+ &=
\frac{1}{\sqrt{2}}
\left(
1,0,0,\zeta ^0,-\tilde{\zeta }_1,-6 \zeta ^1,0,0
\right)
\\
Y_- &=
\frac{1}{\sqrt{2}}
\left(
\tau _1^2-\tau _2^2,2 \tau _1 \tau _2,-3 \zeta ^1,\frac{2 \tilde{\zeta }_1}{3},0,3 \tilde{\zeta }_0,0,0
\right)
\end{aligned}
\end{equation}
Here we display 9 of the fourteen Killing vectors. The others are more complicated and can be found in appendix \ref{KillingAppendix}.

Using equation \eqref{prepotentialFormula} we can then calculate the Killing prepotentials associated with the Killing vectors
\begin{equation}
\begin{aligned}
\Et &=
\left(
\begin{tabular}{c}
$0$ 		\\
$0$	 	\\
$-\frac{1}{4} e^{-2 U}$
\end{tabular}
\right)
,\qquad
\Ht =\frac{e^{-U}}{2\sqrt{2}\tau_2^{3/2}}
\left(
\begin{tabular}{c}
$\tau _2 \left(\tilde{\zeta }_1-3 \tau _1^2 \zeta ^0+\tau _2^2 \zeta ^0+6 \tau _1 \zeta ^1\right)$	 \\
$-\tau _1 \left(\tilde{\zeta }_1+3 \tau _2^2 \zeta ^0\right)-\tilde{\zeta }_0+\tau _1^3 \zeta ^0-3 \tau _1^2 \zeta ^1+3 \tau _2^2 \zeta ^1$	 \\
$\sqrt{2} \sigma  \tau _2^{3/2} e^{-U}$
\end{tabular}
\right)
\\
\Et_{p^0} &=
\frac{e^{-U}}{2\sqrt{2} \tau_2^{3/2}}
\left(
\begin{tabular}{c}
$0$	 \\
$1$	 \\
$\sqrt{2} \tau_2^{3/2} \zeta ^0 e^{-U}$
\end{tabular}
\right)
,\qquad
\Et_{q_0} =
\frac{e^{-U}}{2\sqrt{2} \tau_2^{3/2}}
\left(
\begin{tabular}{c}
$\tau _2^3-3 \tau _1^2 \tau _2$	 \\
$\tau _1^3-3 \tau _1 \tau _2^2$	 \\
$\sqrt{2} \tau _2^{3/2} \tilde{\zeta }_0 e^{-U}$
\end{tabular}
\right)
\\
\Et_{p^1} &=
\frac{3 e^{-U}}{2\sqrt{6} \tau_2^{3/2} }
\left(
\begin{tabular}{c}
$-\tau _2$	 \\
$\tau _1$	 \\
$\sqrt{2} \tau _2^{3/2}  \zeta ^1  e^{-U}$
\end{tabular}
\right)
,\qquad
\Et_{q_1} =
\frac{ e^{-U}}{2\sqrt{6} \tau_2^{3/2} }
\left(
\begin{tabular}{c}
$6 \tau _1 \tau _2$	 \\
$3 \left(\tau _2^2-\tau _1^2\right)$	 \\
$\sqrt{2} \tau _2^{3/2}\tilde{\zeta }_1   e^{-U} $
\end{tabular}
\right)
\\
\Yt_0 &=
-\frac{e^{-U}}{4 \sqrt{2} \tau_2^{3/2}}
\left(
\begin{tabular}{c}
$\tau _2 \left(-\tilde{\zeta }_1-9 \tau _1^2 \zeta ^0+3 \tau _2^2 \zeta ^0+6 \tau _1 \zeta ^1\right)$	 \\
$3 \left(\tilde{\zeta }_0+\tau _1^3 \zeta ^0-3 \tau _1 \tau _2^2 \zeta ^0+\tau _2^2 \zeta ^1-\tau _1^2 \zeta ^1\right)+\tau _1 \tilde{\zeta }_1$	 \\
$\sqrt{2\tau _2} \left(3 \tau _2 \tilde{\zeta }_0 \zeta ^0+\tau _2 \tilde{\zeta }_1 \zeta ^1+3 \tau _1 e^{2 U}\right)e^{-U} $
\end{tabular}
\right)
\\
\Yt_{+} &=
\frac{e^{-U}}{4 \sqrt{2} \tau_2^{3/2}}
\left(
\begin{tabular}{c}
$6 \sqrt{2} \tau _2 \left(\zeta ^1-\tau _1 \zeta ^0\right)$	 \\
$-\sqrt{2} \left(\tilde{\zeta }_1-3 \tau _1^2 \zeta ^0+3 \tau _2^2 \zeta ^0+6 \tau _1 \zeta ^1\right)$	 \\
$\sqrt{\tau _2} \left(3 e^{2 U}-2 \tau _2 \left(\tilde{\zeta }_1 \zeta ^0+3 ( \zeta ^1)^2 \right)\right)e^{-U} $
\end{tabular}
\right)
\\
\Yt_{-} &=
\frac{e^{-U}}{4 \sqrt{2} \tau_2^{3/2}}
\left(
\begin{tabular}{c}
$- \sqrt{2} \tau _2 \left(4 \tau _1 \tilde{\zeta }_1+3 \tilde{\zeta }_0+9 \tau _1^2 \zeta ^1-3 \tau _2^2 \zeta ^1\right)$	 \\
$ \sqrt{2} \left(3 \tau _1 \left(\tilde{\zeta }_0+\tau _1^2 \zeta ^1-3 \tau _2^2 \zeta ^1\right)+2 \left(\tau _1-\tau _2\right) \left(\tau _1+\tau _2\right) \tilde{\zeta }_1\right)$	 \\
$\tau _2^{3/2} \left(
3\frac{\tau _1^2  + \tau _2^2}{\tau _2}
+2 e^{-2 U} \left(3 \tilde{\zeta }_0 \zeta ^1-  \frac{1}{3} \tilde{\zeta }_1^2\right)\right)e^U $
\end{tabular}
\right)
\end{aligned}
\end{equation}

\subsection{The gauging}

On the supergravity side, the R-symmetry group of the undeformed $\mN=8$ theory is $USp(\mN)$, whose maximal subgroup is
\begin{equation}
USp(6) \times USp(2)  \subset USp(8)
\end{equation}
The deformed theory preserves the $USp(2) \cong SU(2)_R$ subgroup, which is the R-symmetry of the resulting $\mN=2$ supergravity.
It is also invariant under $SO(3)$, which is embedded inside $USp(6)$ in the following way
\begin{equation}
SO(3) \times SU(2)_F \subset USp(6)
\end{equation}
namely, $SO(3)$ commutes with an $SU(2)_F$ inside $USp(6)$. The holonomy $\mH$ is the part in $USp(8)$ that commutes with $SO(3)$. It is therefore given by
\begin{equation}
\mH = SU(2)_R \times SU(2)_F
\end{equation}
The generators of $SU(2)_R$ are given by $J_i$ and those of $SU(2)_F$ are $S_i$.

The $U(1)_R$ symmetry of the field theory corresponds to the following combination of $U(1)$ factors inside $\mH$
\begin{equation}
R = S_3 - 3 J_3
\end{equation}
The reason is that the $SU(2)_F$ generators transform under $U(1)_R$ with charge $+1$ and those of $SU(2)_R$ transform with charge $-3$.
The isometry $R$ is the one we want to gauge using the graviphoton. Using \eqref{J-generators} and \eqref{S-generators}, we find that
\begin{equation}
R= - \sqrt{2} \left( Y_+ + Y_-  \right)
\end{equation}

In the absence of vector multiplets there is a description in terms of a superpotential
\begin{equation}
W^2= \frac{2}{3} P^r P^r
\end{equation}
In our case, the prepotential associated with the gauged isometry is given by
\begin{equation}
P^r = \sqrt{\frac{2}{3}} \tilde{R}^r = -  \frac{2}{\sqrt{3}}  (\tilde{Y}_+^r + \tilde{Y}_-^r)
\end{equation}
The superpotential is therefore given by
\begin{equation}\label{superPot2}
\begin{aligned}
 W^2 &=\left[\frac{ \left(\tau _1^2+\tau _2^2+1\right)}{2 \tau _2}
 - \frac{1}{9}  e^{-2 U} \left(
( 3 \zeta ^0 +\tilde{\zeta }_1)\tilde{\zeta }_1 
 - 9 \left(\tilde{\zeta }_0-\zeta ^1\right) \zeta ^1
 \right)\right]^2
\\ &+
\frac{e^{-2 U} }{18 \tau _2}\left[2 \tau _1 \left(  3 \zeta ^0+2 \tilde{\zeta }_1\right)+3( \tilde{\zeta }_0-2\zeta ^1)  +  3 (3 \tau _1^2 - \tau _2^2)\zeta ^1 \right]^2
\\ &+
\frac{e^{-2 U}}{18 \tau _2^3}
 \left[   -  (\tau _1^2 -  \tau _2^2) \left(3 \zeta ^0+2 \tilde{\zeta }_1\right)
-3 \tau _1(\tilde{\zeta }_0-2\zeta ^1)
-3 \tau_1 (   \tau _1^2   - 3 \tau _2^2 ) \zeta ^1
 +\tilde{\zeta }_1
 \right]^2
\end{aligned}
\end{equation}
The superpotential \eqref{superPot2} is one of our main results in this paper.

We can now evaluate the the potential using \eqref{bogo}.
The full expression is quite lengthy and we will not include it here, but it is straightforward to derive it.
The expansion of the potential around the maximally supersymmetric fixed point is
\begin{equation}
V = -6 + \frac{1}{2} \left[
-4 \left( \sqrt{\frac{3}{2}}\tau_1 \right)^2 -4 \left( \sqrt{\frac{3}{2}} (\tau_2-1)\right) ^2-3(\zeta^0)^2 -3(\sqrt{3}\zeta^1)^2 -3 (\zetat_0)^2-3 (\frac{1}{\sqrt{3}}\zetat_1)^2
\right]
+\dots
\end{equation}
where the canonically normalized fields are
\begin{equation}
\left( \sqrt{\frac{3}{2}}\tau_1,  \sqrt{\frac{3}{2}} (\tau_2-1),  \zeta^0,  \sqrt{3}\zeta^1, \zetat_0,  \frac{1}{\sqrt{3}}\zetat_1\right)
\end{equation}
The masses of the scalars are therefore given by
\begin{equation}
\begin{aligned}
m_{\tau_1} ^2 &= m_{\tau_2-1}^2 = -4 \\
m_{\zeta^0} ^2 &= m_{\zeta^1}^2=m_{\zetat_0} ^2 = m_{\zetat_1}^2 = -3 \\
m_{U} ^2 &= m_{\sigma}^2 =0
\end{aligned}
\end{equation}
$\tau_1$ and $\tau_2$ therefore correspond to dimension $\D= 2$ operators, which are the two scalar deformations.
$\zeta^0,\zeta^1,\zetat_0$ and $\zetat_1$ correspond to dimension $\D= 3$ operators, which are the two fermion bilinears and the two gaugino bilinears.
$U$ and $\sigma$ correspond to the complex gauge coupling, which is a marginal deformation $\D=4$.
Note that $\sigma$ does not appear at all in the potential and is therefore a non-linear realization of one flat direction (while $U$ is only a linear realization of the second flat direction).

\subsection{R-symmetry basis}

The R-symmetry generator is given by
\begin{equation}
\begin{aligned}
R = 
&- \left(  1+ \tau _1^2-\tau _2^2  \right)   \pa_{\tau _1}
-2   \tau _1 \tau _2   \pa_{\tau _2}  \\
& +3  \zeta ^1  \pa_{\zeta^0}
+\zetat_1 \pa_{\zetat_0}
-  \left(   \zeta ^0 +  \frac{2}{3}  \zetat _1    \right) \pa_{   \zeta ^1  }  
-3   \left( \zetat_0        -2 \zeta ^1     \right)    \pa_{\zetat_1} 
\end{aligned}
\end{equation}
We would like to find eigenstates of the R-symmetry generator - namely, the combinations of fields that are transformed by a phase under $R$.
These are given by
\begin{equation}\label{R-symmetry-Var}
\begin{aligned}
z_1 &=	 +(\zeta_1-\frac{1}{3} \zetat_0)	- \frac{i}{3}  (\zeta_0+\zetat_1) 	\\
\zb_1 &=	 -(\zeta_1-\frac{1}{3} \zetat_0)	- \frac{i}{3}  (\zeta_0+\zetat_1) 	\\
z_2 &=	-(\zetat_0+\zeta_1)  - i (\zeta_0   -   \frac{1}{3} \zetat_1)	\\
\zb_2 &=	-(\zetat_0+\zeta_1)  + i (\zeta_0   -   \frac{1}{3} \zetat_1)	\\
x &= \frac{i+\taub}{i-\taub},
\qquad
\xb = \frac{i-\tau}{i+\tau}
\end{aligned}
\end{equation}
Note that in Lorentzian signature $\zb_1 = - z_1^*$ and $\zb_2=+z_2^*$, but in the Euclidean theory these fields are independent.
The inverse relations are given by
\begin{equation}
\begin{aligned}
\zeta_0 &= +\frac{3i}{8} \left[  (  z_1 +\zb_1   )  + ( z_2 -\zb_2)   \right]		\\
\zetat_0 &= -\frac{3}{8}  \left[  (  z_1 -\zb_1  )  + ( z_2 +\zb_2)   \right]		\\
\zeta_1 &= + \frac{1}{8}   \left[    3  (  z_1 -\zb_1  )   -   (  z_2 +\zb_2 )    \right]  \\
\zetat_1 &=  +\frac{3i}{8}   \left[    3 (  z_1 +  \zb_1  )   -   (  z_2 - \zb_2 )    \right]  \\
\taub &= - i \frac{1-x}{1+x} ,\qquad  \tau = i \frac{1-\xb}{1+\xb}
\end{aligned}
\end{equation}
Using the variables \eqref{R-symmetry-Var} the R-symmetry generator takes the form
\begin{equation}\label{R-generator2}
R = 
i \left(   2x \pa_x   -3 z_1  \pa_{z_1}+z_2  \pa_{z_2}   \right)
-i \left(   2\xb \pa_{\xb}   -3 \zb_1  \pa_{\zb_1}+\zb_2  \pa_{\zb_2}   \right)
\end{equation}
To get the correct R-charge we need to multiply by $-\frac{2}{3}$
\begin{equation}
-\frac{2}{3} R = 
i \left(   r_x \,x \pa_x   +r_{z_1} z_1  \pa_{z_1}+ r_{z_2} z_2  \pa_{z_2}   \right)
-i \left(   r_{\xb} \,\xb \pa_{\xb}   +r_{\zb_1} \zb_1  \pa_{\zb_1}+ r_{\zb_2} \zb_2  \pa_{\zb_2}   \right)
\end{equation}
where $r_x,r_{\xb},r_{z_1},r_{\zb_1},r_{z_2}$ and $r_{\zb_2}$ are the R-charges of the fields.
We then see that the bulk fields $z_1$ corresponds to the gaugino bilinear, which transforms as $\bs[1,- 6]$ under $SO(3)\times U(1)_R$.
The field $z_2$ corresponds to the fermion bilinear, which transforms as $\bs[1, +2]$.
The field $x$ corresponds to the scalar deformation, which transforms as $\bs[1, +4]$.
The formally-conjugated fields transform with the opposite R-charges.
$U$ and $\sigma$ are inert under R-symmetry transformations and therefore do not appear in \eqref{R-generator2}.

Let us summarize the duality between the scalar fields in the bulk and the operators in the field theory.
The four massive operators in \eqref{SO(3)Lagrangian} are dual to the following bulk fields
\begin{equation}
\begin{aligned}
x \qquad & \longleftrightarrow  \qquad \mO_{x} = \text{Tr}\left(Z^2\right)   \\
\xb \qquad & \longleftrightarrow \qquad  \mO_{\xb} = \text{Tr} \left(  \Zb^2 \right)  \\
z_2 \qquad & \longleftrightarrow \qquad  \mO_{z_2} = \text{Tr} \left( \chi  \chi  \right) \\
\zb_2 \qquad & \longleftrightarrow  \qquad  \mO_{\zb_2} =  \text{Tr}  \left( \chib \chib  \right)
\end{aligned}
\end{equation}
In addition to the massive operators, the spectrum of the theory also contains the gauge kinetic term, the $\theta$-term and left-handed and right-handed gaugino bilinears
\begin{equation}
\begin{aligned}
z_1 \qquad & \longleftrightarrow\qquad  \mO_{z_1} = \text{Tr} \left( \lam \lam \right)  \\
\zb_1 \qquad & \longleftrightarrow \qquad  \mO_{\zb_2} =  \text{Tr}  \left( \lamb \lamb \right)   \\
U \qquad & \longleftrightarrow\qquad   \mO_{U} = \text{Tr} \left( F _{\mu\nu} F^{\mu\nu} \right) \\
\sigma \qquad & \longleftrightarrow\qquad   \mO_{\sigma} = \text{Tr} \left( \ep^{\mu\nu\rho\sigma} F_{\mu\nu}F_{\rho\sigma} \right)
\end{aligned}
\end{equation}

The superpotential in the R-symmetry basis takes the form
\begin{equation}\label{RsuperPotential}
\begin{aligned}
W^2  &=
\left(    \frac{1+  |x|^2   }{1-  |x|^2   }    \right)^2
   -\frac{\mathcal{T}  }{4 (1- |x|^2)^3}   e^{-2U}
   +\frac{1}{64} (9 |z_1|^2 +  |z_2|^2 )^2    e^{-4U}
\end{aligned}
\end{equation}
where
\begin{equation}
\begin{aligned}
\mathcal{T} &=
 6 \xb^2 \zb_1 (3 \xb \zb_1-z_2)+\zb_2 (6 \xb \zb_1-z_2)+9 |z_1|^2   
 \\  & 
-x \left(6 \xb^3 \zb_1 z_2-2 \xb^2 z_2^2-9 \xb |z_1|^2 +\xb | z_2|^2  +6 z_1 z_2\right)
 \\  & 
+x^2 \left(6 \xb^3 \zb_1 \zb_2+\xb^2 (9 | z_1|^2 - |z_2|^2 )+2 \xb \zb_2^2+6 z_1 \zb_2\right)
 \\  & 
+x^3 \left(  -  \xb^3 |z_2|^2  +3 \xb z_1 \left(3 \xb^2 \zb_1-2 \xb z_2+2 \zb_2\right)+18 z_1^2\right)
\end{aligned}
\end{equation}
The notations are $|x|^2\equiv x\xb,|z_1|^2\equiv z_1 \zb_1,|z_2|^2\equiv z_2 \zb_2$. In Euclidean signature the fields are not complex conjugates of each other and therefore those expressions do not represent the absolute values of the fields.

The metric on the scalar manifold in the R-symmetry basis takes the form
\begin{equation}\label{Rmetric}
ds^2 = 2(u \ubar  +v \vbar  +e^1 \ebar ^1  + E_1 \Ebar _1 )
\end{equation}
with the vielbeins
\begin{equation}
\begin{aligned}
u &=		-\frac{3 e^{-U}}{2 \sqrt{2}\left(  1-x \xb  \right)^{3/2}}	
\left(
dz_1+\xb d \zb_2+\xb^2 dz_2 -\xb^3 d\zb_1
\right)
	\\
v &=		dU  +\frac{1}{16} e^{-2U}
\left(
8i d \sigma -9 \zb_1 dz_1 +9 z_1 d\zb_1 -3 z_2 d\zb_2 +3 \zb_2 dz_2
\right)
		\\
e_1&=	  \frac{\sqrt{3}}{1-x \xb} d \xb	\\
E_1&=	\sqrt{\frac{3}{2 (1- x \xb)^3}} e^{-U}\left(
3 x dz_1 -3 \xb^2 d\zb_1
+\xb (2+ x\xb) dz_2
+(1+2 x \xb) d\zb_2
\right)
	\\
\end{aligned}
\end{equation}

\section{Solutions of the coset model}\label{SolutionsSec}

In this section we study different solutions of the theory we have derived in the previous section and find analytical solutions for them.

\subsection{All tilded fields are set to zero}

First we study the truncation $\xb=\zb_1=\zb_2=0$ that correspond to setting $\mt=0$ in the field theory.
The superpotential then takes a simple form
\begin{equation}
W = \sqrt{ 1+ \frac{3}{2} x \, z_1 \left(  z_2 - 3x^2 z_1   \right) e^{-2U}}
\end{equation}

This is not a consistent truncation of the theory.
To explain that let us focus on the fields $x,\xb$, for example. The metric on the scalar manifold is complicated, but the $x$- and $\xb$- components are given by
\begin{equation}
ds^2 = \frac{6}{(1-|x|^2)^2} dx d\xb+\dots
\end{equation}
where the dots refer to the rest of the components.

Now, let us look at the BPS equations
\begin{equation}
q^X \,\! ' \sim \mp \frac{3}{L} g^{XY} \pa_Y W
\end{equation}
Since the inverse metric $g^{XY}$ mixes between $x$ and $\xb$ we will get a non-trivial equation for $\xb$
\begin{equation}
\xb ' \sim \mp \frac{3}{L} g^{\xb x} \pa_{x} W \neq 0
\end{equation}
that will not be consistent with $\xb=0$.

A similar issue occurs with the other fields and therefore this truncation is not-consistent.

\subsection{No axion-dilaton and no gauginos}

We now wish to set the bulk fields dual to the axion-dilaton, $U$ and $\sigma$, and the gauginos $z_1,\zb_1$ to zero
\begin{equation}
U=\sigma=z_1=\zb_1=0
\end{equation}
The superpotential then reduces to
\begin{equation}
\begin{aligned}
W^2= & 
+\frac{(1+|x|^2)(1-|x|^4)}{(1-|x|^2)^3}
+\left( \frac{(1+|x|^2)(1+|x|^4)}{(1-|x|^2)^3} 
+\frac{|z_2|^2}{16}  \right) \frac{ |z_2|^2}{4}\\
&-\frac{|x|^2}{2(1-|x|^2)^3}(x\zb_2^2+\xb z_2^2)
\end{aligned}
\end{equation}
and the potential is given by
\begin{equation}\label{xzPotential}
\begin{aligned}
V=& 
-6 \frac{1+|x|^4}{(1-|x|^2)^2}
+\frac{3}{8}\left(
\frac{|z_2|^2}{8} - \frac{(3-|x|^2)(1-3|x|^2)(1+|x|^2)}{(1-|x|^2)^3}
\right) |z_2|^2 \\
& -\frac{3}{4}\frac{1+|x|^4}{(1-|x|^2)^3}(x\zb_2^2+\xb z_2^2)
\end{aligned}
\end{equation}
The metric on the scalar manifold
\begin{equation}
\begin{aligned}
ds^2 & =
\frac{6 dx d\xb}{(1-|x|^2)^2}
+ \frac{3}{64} \frac{16(1+|x|^2)(1+10|x|^2+|x|^4)+3(1-|x|^2)^3 |z_2|^2}{(1-|x|^2)^3} dz_2 d\zb_2
 \\
 &
-\frac{3}{2}\left[\left( 
\frac{3 \zb_2^2}{64}
-\frac{\xb(1+4|x|^2+|x|^4)}{(1-|x|^2)^3}
 \right) dz_2^2
 +\left( 
\frac{3 z_2^2}{64}
-\frac{x(1+4|x|^2+|x|^4)}{(1-|x|^2)^3}
 \right) d\zb_2^2
 \right]
\end{aligned}
\end{equation}

\subsection{Solutions with no back-reaction}\label{NoBackSec}

We now keep only the fields $x$ and $z_2$ and set all the rest to zero ($\xb=\zb_2=z_1=\zb_1=U=\sigma=0$).
Let us remind that the bulk field $x$ is dual to the coupling to the sphere which is proportional to $i\frac{m}{a}$ and $z_2$ is dual to the fermion bilinear which is proportional to $m$.
It turns out that the kinetic term \eqref{Rmetric} vanishes in this case. The superpotential \eqref{RsuperPotential} is trivial
\begin{equation}
W=1
\end{equation}
and therefore the solution for the metric equation of motion in \eqref{EOMs} is the hyperbolic space
\begin{equation}
e^A = \sinh r
\end{equation}

The equations of motion for the fields $z_2$ and $x$ are
\begin{equation}
\begin{aligned}
z_2''(r) +4 A' (r) z_2'(r) +3 z_2 (r) &=0\\
x''(r) +4 A' (r) x'(r) +4 x (r) &=-\frac{1}{4} z_2(r)^2
\end{aligned}
\end{equation}
For which the solution is
\begin{equation}\label{solEOMS}
\begin{aligned}
z_2(r) &=    \frac{1}{4 \sinh(r)^3}  \left[  a_1 \left(  \sinh(2r) - 2  r \right)  +4 a_2  \right]  \\
x(r) &=  \frac{1}{2\sinh(r)^2}  \left[  b_1 \left(  r \coth(r)-1 \right)  +2b_2 \coth(r)     \right] + \frac{a_1 r (4a_2 -a_1 r)}{64\sinh(r)^2}  \\
&- \frac{\coth(r) }{64\sinh(r)^2}  \left[ 
  4a_1 a_2 + (4 a_2^2 -a_1 ^2 )  r +(a_1 r -2 a_2)^2 \coth(r)  
\right]
\end{aligned}
\end{equation}
Imposing regularity at $r=0$ set the coefficients $a_2=b_2=0$.
Supersymmetry further fixes a relation between $a_1$ and $b_1$, as we will explain in the next section.

The Fefferman-Graham expansion of the solution near the boundary $r \rightarrow \infty$  is
\begin{equation}\label{FGexpansion}
\begin{aligned}
z_2(r) &=  a_1 e^{-r }  +a_1 (3 - 4 r)  e^{-3r} +\dots  	  \\
x(r) &=  \left(  -2 b_1 +(2b_1 +\frac{1}{16}a_1^2 )r -\frac{1}{8} a_1^2 r^2   \right) e^{-2r}  +\dots
\end{aligned}
\end{equation}
corresponding to operators of dimensions $\Delta_{\mO_{z_2}}=3$ and $\Delta_{\mO_{x}}=2$, respectively.
$a_1$ is the source for the scalar operator $\mO_{z_2}$ and $b_1$ is the source for the operator $\mO_{x}$ .

\subsection{Flat spacetime truncation with a dilaton and a gaugino condensate}

If we set both $x$ and $\xb$ to zero we get some nice and simple truncations with analytical solutions.
However, $x$ and $\xb$ correspond to the scalar deformations which encode the coupling to $S^4$.
Therefore, if we set them both to zero we truncate to flat spacetime solutions.

Let us look for example at the truncation
\begin{equation}
x=\xb=z_2=\zb_2=\sigma=0
\end{equation}
The superpotential is then given by
\begin{equation}
W=1-  \frac{9}{8} |z_1|^2   e^{-2U}
\end{equation}
The solution for the scalars is
\begin{equation}
\begin{aligned} 
z_1(r) &= z_0 \,  e^{-3r} \\
\zb_1(r) &= \zb_0  \, e^{-3r} \\
U(r) &= \frac{1}{2} \ln \left(	   1-\frac{9}{8} |z_0|^2 e^{-6r}	\right) 
\end{aligned}
\end{equation}
For flat space solutions the equation of motion of the wrap factor is $A'(r)=W$, for which the solution is 
\begin{equation}
A(r) = r+ \frac{1}{6} \ln \left( 	1-\frac{9}{8} |z_0|^2 e^{-6r}	\right)
\end{equation}
This solution is singular.
The singularity is at $r=\frac{1}{6} \ln(\frac{9}{8}|z_0|^{2})$, where the argument of the log term vanishes

A similar solution was found in \cite{Ceresole:2001wi} for $\mN=2$ supergravity coupled to one hypermultiplet.

\section{The Free Energy}\label{FreeEnergySec}

One of the main purposes of this paper is to evaluate the free energy for the $\mN=1^*$ theory on $S^4$ using the analytical solution that we found in section \ref{NoBackSec}.

To evaluate the free energy we need to compute the one-point functions using the procedure of holographic renormalization.
The on-shell bulk action (supplemented by the Gibbons-Hawking term) is divergent and has to be regularized using infinite counterterms.
In addition, as explained in \cite{Bobev:2016nua,Bobev:2013cja,Freedman:2013ryh}, in order to preserve supersymmetry we also need to add finite counterterms.
The renormalized on-shell action is given by
\begin{equation}
S_{ren} = S_{5D} + S_{GH} +S_{ct}+ S_W
\end{equation}
where $S_{5D}$ is the bulk action, $S_{GH}$ is the Gibbons-Hawking term, $S_{ct}$ is the infinite counterterm action and $S_W$ is the finite counterterm.

The bulk action is
\begin{equation}
S_{5D} = \frac{1}{8\pi G_5} \int d^5 x \sqrt{g} 
\left(
-\frac{1}{2} R -\frac{1}{2} g_{XY} \pa_{\mu} q^X \pa^{\mu} q^Y - \frac{1}{L^2} V
\right)
\end{equation}
The potential \eqref{xzPotential} can be expanded around the maximally supersymmetric fixed point
\begin{equation}\label{xzPotentialExp}
V=-6
-\frac{3}{8} \left(  32 |x|^2 -9 |z_1|^2 +3 |z_2|^2 \right)
-\frac{3}{4} \xb z_2^2
+\dots
\end{equation}
where $\dots$ represent terms that will vanish in the limit where the cutoff is taken to infinity.

The infinite counterterm action is given by
\begin{equation}
\begin{aligned}\label{infiniteCT}
S_{ct} &=-\frac{1}{8 \pi G_5} 
\int _{\rho=\ep} d^4x \sqrt{\gamma} \left[
\frac{3}{2} +\frac{1}{8} R[\gamma]
+\frac{3}{8} |z_2|^2 
+ 6 \left(1+\frac{1}{\log \rho}  \right) |x|^2
\right. \\ & \left.
-\log \rho \left(
\frac{1}{32} \left( R[\gamma]_{ij} R[\gamma]^{ij}  -\frac{1}{3} R[\gamma]^2 \right)
-\frac{1}{24} R[\gamma] \left( \frac{3}{8} |z_2|^2 \right)
+\frac{3}{16} \xb z_2^2
\right)
\right]
\end{aligned}
\end{equation}
We have define a radial coordinate $\rho = e^{-2r}$ and $\ep$ is the cutoff. $\gamma$ is the induced metric on the boundary.
Most of the terms in \eqref{infiniteCT} are the canonical counterterms for bulk fields dual to operators of dimension $\Delta_{z_2}=3$ and $\Delta_x=2$ on $S^4$ (see \cite{Bobev:2016nua}).
The only exception is the last term, proportional to $\xb z_2^2$, which is a result of the last interaction term in the expansion of the potential \eqref{xzPotentialExp}.

Finally, we also have to had a finite counterterm. As explained in \cite{Bobev:2016nua,Bobev:2013cja,Freedman:2013ryh}, to preserve supersymmetry we need to add to the action the finite part of the following term
\begin{equation}
S_W = -\frac{3}{8\pi G_5}  \int _{\pa}  d^4x \sqrt{\gamma} \, W
\end{equation}
where $W$ is the superpotential.
$S_W$ contains finite and infinite terms, but the infinite ones were already included in $S_{ct}$, so we only need to consider finite contributions.

The renormalized one-point functions for operators $\mO_{\D}$ of dimension $\D=2,3,4$ dual to the bulk fields $\Phi_{\D}$ are given by
\begin{equation}
\begin{aligned}
\left< \mO_2 \right> &= \lim_{\ep \rightarrow 0} \frac{\log \ep}{\ep}  \frac{1}{\sqrt{\gamma}} \frac{\delta S_{ren}}{\delta \Phi_2}
= \lim_{r \rightarrow \infty} \left(  -2r e^{2r}  \frac{1}{\sqrt{\gamma}} \frac{\delta S_{ren}}{\delta \Phi_2} \right) \\
\left< \mO_3 \right> &= \lim_{\ep \rightarrow 0}   \frac{1}{\ep ^{3/2}}  \frac{1}{\sqrt{\gamma}} \frac{\delta S_{ren}}{\delta \Phi_3} 
= \lim_{r \rightarrow \infty} \left(  e^{3r}  \frac{1}{\sqrt{\gamma}} \frac{\delta S_{ren}}{\delta \Phi_3}  \right)  \\
\left< \mO_4 \right> &= \lim_{\ep \rightarrow 0} \frac{1}{\ep ^2}  \frac{1}{\sqrt{\gamma}} \frac{\delta S_{ren}}{\delta \Phi_4}
= \lim_{r \rightarrow \infty}  \left(  e^{4r}  \frac{1}{\sqrt{\gamma}} \frac{\delta S_{ren}}{\delta \Phi_4}  \right)
\end{aligned}
\end{equation}

The contribution to the renormalized one-point functions from the bulk action is
\begin{equation}
\frac{1}{\sqrt{\gamma}}\frac{\delta S_{5D}}{\delta q^X} = 
-\frac{1}{8\pi G_5} \left(  g_{XY} \pa_{r} q^Y  \right)
\end{equation}
For the solution in section \ref{NoBackSec} the only non-zero components of the above expression are
\begin{equation}
\begin{aligned}
\frac{1}{\sqrt{\gamma}}\frac{\delta S_{5D}}{\delta \xb} &=   - \frac{1}{8\pi G_5} \left(    3  \pa_{r} x   \right) \\
\frac{1}{\sqrt{\gamma}}\frac{\delta S_{5D}}{\delta \zb_2} &=- \frac{1}{8\pi G_5} \left(     \frac{3}{8}  \pa_{r} z_2  \right)
\end{aligned}
\end{equation}
The contribution from the infinite counterterms is
\begin{equation}
\begin{aligned}
\frac{1}{\sqrt{\gamma}}\frac{\delta S_{ct}}{\delta \xb} &= - \frac{1}{8\pi G_5} \left[  6 \left(  1-\frac{1}{2r} \right) x +r \frac{3}{8} z_2^2  \right]  \\
\frac{1}{\sqrt{\gamma}}\frac{\delta S_{ct}}{\delta \zb_2} &=-  \frac{1}{8\pi G_5} \left[   \frac{3}{8} \left( 1-8 r e^{-2r} \right) z_2     \right]
\end{aligned}
\end{equation}
where, again, derivatives with respect to any other fields are zero.
The superpotential contribution is
\begin{equation}
\begin{aligned}
\frac{1}{\sqrt{\gamma}}\frac{\delta S_{W}}{\delta \xb} &= -  \frac{1}{8\pi G_5} \left(  6x \right) \\
\frac{1}{\sqrt{\gamma}}\frac{\delta S_{W}}{\delta \zb_2} &= -  \frac{1}{8\pi G_5} \left(  \frac{3}{8} z_2 \right)
\end{aligned}
\end{equation}
Note that the superpotential contribution does not contain any finite pieces! The infinite contributions of $S_W$ were already taken into account in $S_{ct}$.

Plugging the solution \eqref{solEOMS} and adding everything together we find the renormalized one-point functions
\begin{equation}
\begin{aligned}
\left< \mO_{\zb_2} \right> &=\frac{15}{4} a_1\\
\left< \mO_{\xb} \right> &=12 b_1  
\end{aligned}
\end{equation}
All other one-point functions vanish.

We can now use the SUSY Ward identities to fix the relation between $a_1$ and $b_1$. Following \cite{Bobev:2016nua}), the SUSY Ward identity for these operators is
\begin{equation}
\sqrt{2}  \left< \mO_{\zb_2} \right>  =   \left< \mO_{\xb} \right>  
\end{equation}
(the normalization is different than the one in \cite{Bobev:2016nua} due to a different normalization of the scalars).
We therefore conclude that a supersymmetric solution should obey
\begin{equation}\label{SUSYsol}
a_1 = \frac{8 \sqrt{2}}{5} b_1
\end{equation}

Finally, as explained in \cite{Freedman:2013ryh}, the free energy for configurations of the form discussed in section \ref{NoBackSec}, is given by the Legendre transform of the action
\begin{equation}\label{Legendre}
F=S_{ren}-\frac{1}{2}\int_{S^4}d^4x
\left[
(a_1+\ab_1)\left( \frac{\delta S_{ren}}{\delta a_1}  +  \frac{\delta S_{ren}}{\delta \ab_1}  \right)
+(b_1+\bbar _1)\left( \frac{\delta S_{ren}}{\delta b_1}  +  \frac{\delta S_{ren}}{\delta \bbar _1}  \right)
\right]
\end{equation}
The contribution of the first term in \eqref{Legendre} is schematically of the form
\begin{equation}
S_{ren} = F_0 + \frac{1}{2}\int_{S^4}d^4x \sqrt{g_0} \left[ a_1   \frac{\delta S_{ren}}{\delta a_1}  + b_1  \frac{\delta S_{ren}}{\delta x}  \right]
\end{equation}
since $a_1$ and $b_1$ are the only sources we turn on. $F_0$ is the contribution from the UV fixed point - $\mN=4$ SYM, and $g_0$ is the metric on $S^4$.
The derivatives of the renormalized action with respect to the sources are related to the one-point functions as follows
\begin{equation}
\begin{aligned}
 \frac{\delta S_{ren}}{\delta \ab_1} &= \frac{\delta S_{ren}}{\delta \zb_2}  \frac{\delta \zb_2}{\delta \ab_1} 
 =\sqrt{g_0} \left< \mO_{\zb_2} \right>,\qquad
  \frac{\delta S_{ren}}{\delta a_1} = \frac{\delta S_{ren}}{\delta z_2}  \frac{\delta z_2}{\delta a_1} 
 =\sqrt{g_0} \left< \mO_{z_2} \right>
  \\
  \frac{\delta S_{ren}}{\delta \bbar_1} &= \frac{\delta S_{ren}}{\delta \xb}  \frac{\delta \xb}{\delta \bbar_1} 
 = - \sqrt{g_0} \left< \mO_{\xb} \right> , \qquad
   \frac{\delta S_{ren}}{\delta b_1} = \frac{\delta S_{ren}}{\delta x}  \frac{\delta x}{\delta b_1} 
 = - \sqrt{g_0} \left< \mO_{x} \right> 
 \end{aligned}
\end{equation}
All other derivatives vanish.
Note that since $\left< \mO_{z_2} \right>$ and $\left< \mO_{x} \right>$ vanish for the solution that we found, the only contribution from the first term in \eqref{Legendre} is the $\mN=4$ SYM part - $F_0$.
We can now evaluate the free energy
\begin{equation}
\begin{aligned}
F &=F_0-\frac{1}{2}\int_{S^4}d^4x \sqrt{g_0}
\left(
a_1 \left< \mO_{\zb_2} \right>
-b_1 \left< \mO_{\xb} \right> 
\right) \\
&= 
F_0 -\frac{1}{2} \text{vol}(S^4)
\left(
a_1 \left< \mO_{\zb_2} \right>
-b_1 \left< \mO_{\xb} \right> 
\right) \\
&= 
F_0 +\frac{ \text{vol}(S^4)}{2}\frac{1}{8 \pi G_5} 
\left(
12 b_1^2-\frac{15}{4} a_1 ^2
\right)
 \end{aligned}
\end{equation}
We use that $\frac{1}{8\pi G_5} = \frac{N^2}{4 \pi^2}$ and that the volume of the 4-sphere with radius 1/2 is $ \text{vol}(S^4) = \frac{1}{2^4} \frac{8 \pi^2}{3} = \frac{\pi^2}{6}$ to get
\begin{equation}
F = F_0 + \frac{N^2}{48}\left(
12 b_1^2-\frac{15}{4} a_1 ^2
\right)
\end{equation}
For the supersymmetric solution \eqref{SUSYsol} we have
\begin{equation}
F = F_0 - N^2 \frac{15}{512} a_1^2
\end{equation}

$a_1$ is proportional to the mass parameter $\mu$. To fix the normalization we recall \cite{Bobev:2016nua} that when the masses are unequal we have
\begin{equation}
\begin{aligned}
\phi_i &= \mu_i e^{-r} +\dots \\
\bar{\phi}_i &= \bar{\mu}_i e^{-r} +\dots 
\end{aligned}
\end{equation}
where $\mu_i=m_i a,\bar{\mu}_i=\mb_i a$ are the dimensionless mass parameters.
In the equal mass case $\mu_i=\mu, \bar{\mu}_i = \bar{\mu}$ the kinetic term is of the form
 \begin{equation}
 \sum_{i=1}^3 \pa \phi_i \pa \bar{\phi}_i
\rightarrow
3\pa \phi \pa \bar{\phi} =\frac{3}{4} \pa z_2 \pa \zb_2
\end{equation}
where the normalization of the $z_2,\zb_2$ variables follows from \eqref{Rmetric}.
We therefore conclude that
\begin{equation}
\begin{aligned}
a_1 &= 2 \mu =2 m a\\
\bar{a}_1 &= 2\bar{\mu} = 2 \mb a
\end{aligned}
\end{equation}
The free energy is therefore given by
\begin{equation}\label{mainResult}
F =
F_0 -  \frac{15}{128} N^2 (ma)^2
\end{equation}
This is our main result.
The expression \eqref{mainResult}, calculated using the gravity dual of the $\mN=1^*$ theory, provides an analytical prediction for its sphere partition function at large 't Hooft coupling in the planar limit.
The free energy of this configuration is quadratic in the mass, and as explained in the introduction, is devoid of unphysical ambiguities.

\section{Concluding remarks and future directions}\label{conclusions}

In this paper we have studied the $SO(3)$ sector of the $\mN=1^*$ mass deformation of $\mN=4$ super Yang-Mills on $S^4$. The gravity dual of this sector is $\mN=2$ supergravity coupled to two hypermultiplets, which is a consistent truncation of the maximally supersymmetric $\mN=8$ supergravity in five dimensions.
The scalar fields in the hypermultiplets span an eight-dimensional quaternionic-\Kahler manifold that is described by the $G_{2,2}/SU(2)\times SU(2)$ coset model.
We have studied the coset model and derived a superpotential for this theory.
Using the superpotential description, we found field configurations in the bulk that feature analytical solutions. We then used these solutions to compute the $S^4$ partition function using the procedure of holographic renormalization, and showed that it is devoid of unphysical ambiguities. An interesting feature of the result \eqref{mainResult} is that it is quadratic in the dimensionless mass parameter $ma$.

The result \eqref{mainResult} provides an analytical prediction for the sphere partition function of the configuration \eqref{configuration} of the $\mN=1^*$ theory at large 't Hooft coupling in the planar limit.
While traditional field theory techniques usually cannot be applied in the strong coupling limit, supersymmetric localization makes it possible in certain cases.
On the four-sphere, however, the localization technique requires the existence of at least $\mN=2$ supersymmetries, and therefore cannot be applied in the case of the $\mN=1^*$ theory.
It will be very interesting if one could develop tools that will allow for the study of quantum field theories with $\mN=1$ sueprsymmetry in the strong coupling regime, and compare with the result \eqref{mainResult}.

The main purpose of this paper was to study the $G_{2,2}/SU(2)\times SU(2)$ coset model and derive analytical results for the $S^4$ partition function of the configuration \eqref{configuration}.
We would like to generalize some of the results that we have derived.
In \cite{Kol:ToAppear1} we extend the analysis and compute the BPS equations for general $\mN=2$ Lorentzian and Euclidean supergravity theories. We also provide a more general derivation of the the holographic renormalization procedure, including both finite and infinite counterterms, that applies for this wide class of theories.

It will be very interesting to use the results and techniques that we developed here to compute other observables, like Wilson loops in $\mN=1$ theories, along the lines of \cite{Russo:2013qaa,Russo:2013kea,Chen-Lin:2015dfa,Chen-Lin:2015xlh}.
It will also be interesting to test holography in other cases where field theory results are available, like $\mN=1$ supersymmetric theories on $S^3\times S^1$ or $S^2 \times T^2$ \cite{Closset:2013sxa,Assel:2014paa,Benini:2015noa,Closset:2016arn,Benini:2016hjo}.
We certainly intend to explore these directions.

\acknowledgments

I would like to thank Silviu S. Pufu for very useful discussions.
UK is supported by the Michigan Center for Theoretical Physics and the Research Corporation for Science Advancement.

\appendix

\section{Indices}

\begin{equation}
\begin{aligned}
\mu &\qquad \qquad 0,\dots,4   &\qquad   &\text{Spacetime indices }  \\
i &\qquad \qquad 1,2   &\qquad  &SU(2)\text{-doublets} \\
r &\qquad \qquad 1,2,3   &\qquad  &SU(2)\text{-triplets} \\
I &\qquad \qquad 0,\dots,n_V   &\qquad  &\text{vectors}  \\
x &\qquad \qquad 1,\dots,n_V   &\qquad  &\text{scalars in vector multiplets} \\
A &\qquad \qquad 1,\dots,2n_H   &\qquad  &\text{symplectic index for hypermultiplets} \\
X &\qquad \qquad 1,\dots,4n_H   &\qquad  &\text{scalars in hypermultiplets} 
\end{aligned}
\end{equation}

\section{Clifford Algebra in 5D}\label{Clifford5D}

The five dimensional gamma matrices $\gamma_m$ where $m=0,\dots,4$ satisfy the Clifford algebra
\begin{equation}
\{\gamma_m,\gamma_n \}  = 2\eta_{mn} = 2 \text{diag} \{ 1,-1,-1,-1,-1 \}
\end{equation}
where $\gamma_m$ with $m=0,\dots,3$ are pure imaginary and $\gamma_4$ is pure real.
We also define
\begin{equation}
  \g5=-i \gamma_4
\end{equation}
Complex conjugation
\begin{eqnarray}
  \gamma_4 ^* &=& \gamma_4 \\
  \gmd^* &=& -\gmd \qquad (\mu\neq4) \\
  \g5 ^* &=& -\g5 \\
  (\gmd\g5)^* &=& -\g5\gmd \\
  (\gmu\g5)^* &=& -\g5\gmu
\end{eqnarray}

\section{$SU(2)$ and $Sp(2n_H)$ structures}\label{su2structure}
Partially based on appendix 20A of \cite{Freedman:2012zz}.

\subsection{The Pauli matrices}

The Pauli matrices
\begin{equation}
\left( \tau^1 \right)_i\,^j =
\left(
\begin{tabular}{cc}
0 & 1 \\
1 & 0 \\
\end{tabular}
\right), 
   \qquad
   \left(\tau^2 \right)_i\,^j =
\left(
\begin{tabular}{cc}
0 & -$i$ \\
$i$ & 0 \\
\end{tabular}
\right), 
   \qquad
      \left(\tau^3 \right)_i\,^j =
\left(
\begin{tabular}{cc}
1 & 0 \\
0 & -1 \\
\end{tabular}
\right).
\end{equation}
Useful identities for the Pauli matrices
\begin{eqnarray}
\nonumber [ \tau^a,\tau^b ] =&2i \ep^{abc} \tau^c,\quad\quad &\operatorname{tr} \tau^a = 0,
\qquad\qquad \operatorname{tr} \tau^a \tau ^b = 2 \delta^{ab}
\\
\{\tau^a,\tau^b\}=&2\delta^{ab} \mathbb{I},\quad\quad &\operatorname{det} \tau^a = -1
\end{eqnarray}
and
\begin{equation}
\tau^a \tau^b = i \ep^{abc} \tau^c  + \delta^{ab} \mathbb{I}
\end{equation}

\subsection{$SU(2)$ indices}
The $SU(2)$ index $i=1,2$ is raised and lowered using the $\epsilon$ symbol
\begin{equation}
\ep_{ij} = \ep^{ij} = 
\left(
\begin{tabular}{cc}
0 & 1 \\
-1 & 0 \\
\end{tabular}
\right),
\qquad\qquad
\ep_{12}=-\ep_{21}= +1
\end{equation}
in the following way
\begin{equation}
A^i = \ep^{ij}A_j, \qquad \qquad A_i=A^j \ep_{ji}
\end{equation}
We than get, for example,
\begin{equation}
A^iB_i = - A_iB^i
\end{equation}

We can raise and lower indices on $\vec{\tau}$
\begin{equation}
\begin{aligned}
(\vec{\tau}) ^{ij} &= \ep^{ik}(\vec{\tau})_k\,^j
=\left(  \tau^3, \; i \mathbb{I} ,\; -\tau^1 \right) = - (\vec{\tau})_{ij}\,^*
\end{aligned}
\end{equation}
With the indices at equal height, $\vec{\tau}$ are symmetric matrices.

\subsection{Decomposition in terms of $SU(2)$ triplets}
Any $SU(2)$ matrix $R_{ij}$ can be decompose in terms of $SU(2)$ triplets $R^r$
\begin{equation}
R_{ij}= i R^r (\tau^r)_{ij}, \qquad \qquad  r=1,2,3
\end{equation}
The inverse relation
\begin{equation}
R^r = \frac{i}{2} R_{ij} (\tau^r)^{ij} = -\frac{i}{2} R_i\,^j (\tau^r)_j\,^i
\end{equation}

For example, we can derive the following identity
\begin{equation}
A_{ij}B^{ij} = 2 A^r B^r \equiv 2 \vec{A} \cdot \vec{B}
\end{equation}

\subsection{$Sp(2n_H)$ structure}

The indices $A,B=1,\dots,2n_H$ describe the fundamental representation of $Sp(2n_H)$. They are raised and lowered using the symplectic matrix $C_{AB}$ which satisfies
\begin{equation}
C_{AB}C^{BC} = \delta_A^C, \qquad C^{AB}=(C_{AB})^*
\end{equation}
By redefinition, this matrix can be brought into the form
\begin{equation}
C_{AB} = \left(
\begin{tabular}{cc}
0 & $\mathbb{I}$ \\
$-\mathbb{I}$ & 0 \\
\end{tabular}
\right)
\end{equation}
In the case $n_H=1$ this structure collapses to that of $SU(2)$.

\subsection{Charge conjugation and reality conditions}

The charge conjugation under the $SU(2)$ and $Sp(2n_H)$ is defined by
\begin{equation}
(R_{iA})^C=\ep_{ij}C_{AB}(f_{jB})^*
\end{equation}
Quantities which are real under charge conjugation (like the vielbeins) satisfy the following reality condition
\begin{equation}
f^X_{iA} = (f^X_{iA})^C = \ep_{ij}C_{AB} (f^X_{jB})^*
\end{equation}

\section{$SU(2,1)$ Killing vectors and Prepotentials}\label{KillingApp}
The eight generators of $SU(2,1)$ can be classified as follows \cite{Behrndt:2000ph}:
\begin{enumerate}
\item The generators of the compact subgroup $SU(2) \times U(1)$.
\item The generators of the non-compact coset $\frac{SU(2,1)}{SU(2)\times U(1)} $.
\end{enumerate}

The generators of the compact subgroup $SU(2) \times U(1)$ are given by the following Killing vectors
\begin{equation}
SU(2) \quad  
\begin{cases} 
k_1 &= \frac{1}{2i} \left[ z_2 \pa_{z_1} +z_1 \pa_{z_2}  -c.c.  \right]\\ 
k_2 &= \frac{1}{2} \left[   -z_2 \pa_{z_1} +z_1 \pa_{z_2}  + c.c.  \right] \\
k_3 &= \frac{1}{2i}  \left[ -z_1 \pa_{z_1} +z_2 \pa_{z_2} -c.c.  \right]
\end{cases}
\end{equation}
\begin{equation}
U(1) \quad  
\begin{cases} 
k_4 = \frac{1}{2i} \left[  z_1 \pa_{z_1} +z_2 \pa_{z_2} - c.c.  \right]
\end{cases}
\end{equation}
The action of the $SU(2)$ subgroup corresponds to "rotations" of the two complex coordinates $z_1,z_2$, and the three generators $\left( k_1,k_2,k_3 \right)$ fulfill the $SU(2)$ algebra $\left[k_m,k_n\right]=i\epsilon_{mnl}k_{l}$.

The generators of the non-compact coset $\frac{SU(2,1)}{SU(2)\times U(1)}$ are given by the following Killing vectors
\begin{equation}
\frac{SU(2,1)}{SU(2)\times U(1)} \quad  
\begin{cases} 
k_5 &= \frac{1}{2} \left[  (-1+z_1^2) \pa_{z_1} +z_1z_2\pa_{z_2} +c.c.  \right] \\ 
k_6 &= \frac{i}{2} \left[  (1+z_1^2) \pa_{z_1} +z_1z_2\pa_{z_2} -c.c.  \right]\\
k_7 &= \frac{1}{2} \left[  -z_1z_2 \pa_{z_1} +(1-z_2^2) \pa_{z_2} +c.c.   \right]  \\
k_8 &=\frac{i}{2} \left[  z_1z_2 \pa_{z_1} +(1+z_2^2) \pa_{z_2} -c.c.   \right] 
\end{cases}
\end{equation}

The full $SU(2,1)$ algebra is given by
\begin{equation}
\left[k_m,k_n\right]=i f_{mnl}k_{l}
\end{equation}
with the structure constants
\begin{equation}
\begin{aligned}
f_{123} &= 1 \\
f_{178} &= f_{156} = f_{268} = f_{275} = f_{358} = f_{367} = -\frac{1}{2} \\
f_{458} &= f_{476} =  \frac{\sqrt{3}}{2}
\end{aligned}
\end{equation}

The Killing prepotentials associated with the Killing vectors can be derived using equation \eqref{prepotentialFormula}. They are most conveniently written using the polar system of coordinates. 
The Killing prepotentials associated with the Killing vectors $\left(k_1,k_2,k_3,k_4\right)$ of the compact subgroup $SU(2)\times U(1)$ are given by \cite{Behrndt:2000ph}
\begin{equation}
\begin{aligned}
SU(2) \quad  
& \begin{cases} 
p_1 &= -
\frac{1}{2\sqrt{1-R^2}}
\left(
\begin{tabular}{c}
$\cos\psi\sin\phi+\cos\theta\sin\psi\cos\phi$ \\
$\sin\psi\sin\phi-\cos\theta\cos\psi\cos\phi$ \\
$-\frac{2-R^2}{2\sqrt{1-R^2}}\sin\theta\cos\phi$
\end{tabular}
\right)\\
p_2 &= -
\frac{1}{2\sqrt{1-R^2}}
\left(
\begin{tabular}{c}
$\cos\psi\cos\phi-\cos\theta\sin\psi\sin\phi$ \\
$\sin\psi\cos\phi+\cos\theta\cos\psi\sin\phi$ \\
$\frac{2-R^2}{2\sqrt{1-R^2}}\sin\theta\sin\phi$
\end{tabular}
\right)\\
p_3 &= -
\frac{1}{2\sqrt{1-R^2}}
\left(
\begin{tabular}{c}
$\sin\psi\sin\theta$ \\
$-\cos\psi\sin\theta$ \\
$\frac{2-R^2}{2\sqrt{1-R^2}}\cos\theta$
\end{tabular}
\right)
\end{cases}
\\
U(1) \quad  
& \begin{cases} 
p_4 &= 
\frac{R^2}{4(1-R^2)}
\left(
\begin{tabular}{c}
0 \\
0 \\
1
\end{tabular}
\right)
\end{cases}
\end{aligned}
\end{equation}

The Killing prepotentials associated with the Killing vectors $\left(k_5,k_6,k_7,k_8\right)$ of the non-compact coset are given by \cite{Behrndt:2000ph}
\begin{equation}
\frac{SU(2,1)}{SU(2)\times U(1)} \quad  
\begin{cases} 
p_5 &= 
\frac{R}{2(1-R^2)}
\left(
\begin{tabular}{c}
$\sqrt{1-R^2} \sin\frac{\theta}{2}\cos\frac{\phi-\psi}{2}$ \\
$-\sqrt{1-R^2} \sin\frac{\theta}{2}\sin\frac{\phi-\psi}{2}$ \\
$\cos\frac{\theta}{2}\sin\frac{\phi+\psi}{2}$
\end{tabular}
\right)\\
p_6 &= -
\frac{R}{2(1-R^2)}
\left(
\begin{tabular}{c}
$\sqrt{1-R^2} \sin\frac{\theta}{2}\sin\frac{\phi-\psi}{2}$ \\
$\sqrt{1-R^2} \sin\frac{\theta}{2}\cos\frac{\phi-\psi}{2}$ \\
$-\cos\frac{\theta}{2}\cos\frac{\phi+\psi}{2}$
\end{tabular}
\right)\\
p_7 &= 
\frac{R}{2(1-R^2)}
\left(
\begin{tabular}{c}
$\sqrt{1-R^2} \cos\frac{\theta}{2}\cos\frac{\phi+\psi}{2}$ \\
$\sqrt{1-R^2} \cos\frac{\theta}{2}\sin\frac{\phi+\psi}{2}$ \\
$\sin\frac{\theta}{2}\sin\frac{\phi-\psi}{2}$
\end{tabular}
\right)\\
p_8 &= -
\frac{R}{2(1-R^2)}
\left(
\begin{tabular}{c}
$\sqrt{1-R^2} \cos\frac{\theta}{2}\sin\frac{\phi+\psi}{2}$ \\
$-\sqrt{1-R^2} \cos\frac{\theta}{2}\cos\frac{\phi+\psi}{2}$ \\
$-\sin\frac{\theta}{2}\cos\frac{\phi-\psi}{2}$
\end{tabular}
\right)
\end{cases}
\end{equation}
We follow the conventions of \cite{Ceresole:2001wi}. In order to translate to the conventions of \cite{Behrndt:2000ph}, one has to multiply the prepotentials by a factor of $-\frac{1}{2}$.

\section{Different system of coordinates for the $SU(2,1)/SU(2)\times U(1)$ coset}

Another system of coordinates which is sometimes being used in the literature is given by
\begin{equation}
\begin{aligned}
z_1 & =   \frac{2 (\theta -i \tau )}{1 +V + \theta ^2 +\tau ^2+i \sigma} \\
z_2 &= -1+\frac{2}{1 +V + \theta ^2 +\tau ^2+i \sigma}
\end{aligned}
\end{equation}
In the system of coordinates $(V,\sigma,\theta,\tau)$ the metric takes the form
\begin{equation}
ds^2 = \frac{dV^2}{2V^2} +
\frac{1}{2V^2} \left(  d\sigma+2 \theta d\tau-2\tau d\theta \right)^2
+\frac{2}{V}\left(  d\tau^2 + d\theta^2  \right)
\end{equation}

\section{$G_{2,2}$ Killing vectors and Prepotentials}\label{KillingAppendix}

The rest of the Killing vectors
\begin{equation}
\begin{aligned}
F_{p^0} &=
\left(
-\frac{2}{3} \left(\tau _1^2 \tilde{\zeta }_1+3 \tau _1 \tilde{\zeta }_0-\tau _2^2 \tilde{\zeta }_1\right),
-\frac{2}{3} \tau _2 \left(2 \tau _1 \tilde{\zeta }_1+3 \tilde{\zeta }_0\right),
\tilde{\zeta }_0 \zeta ^0+\tilde{\zeta }_1 \zeta ^1-\sigma -\frac{\tau _1^3 e^{2 U}}{\tau _2^3},
\right. \\ & \left. 
-\frac{2 \tilde{\zeta }_1^2}{9}-\frac{\left(\tau _1^2+\tau _2^2\right) \tau _1^2 e^{2 U}}{\tau _2^3},
-2 \tilde{\zeta }_0^2-\frac{\left(\tau _1^2+\tau _2^2\right){}^3 e^{2 U}}{\tau _2^3},
\frac{3 \tau _1 \left(\tau _1^2+\tau _2^2\right){}^2 e^{2 U}}{\tau _2^3}-2 \tilde{\zeta }_0 \tilde{\zeta }_1,
\right. \\ & \left. 
-\tilde{\zeta }_0 ,
-\tilde{\zeta }_0^2 \zeta ^0-\tilde{\zeta }_0 \tilde{\zeta }_1 \zeta ^1-\sigma  \tilde{\zeta }_0+\frac{2 \tilde{\zeta }_1^3}{27}
\right. \\ & \left. +
\frac{e^{2 U} \left(\tau _1 \left(\tau _1 \left(\tau _1 \tilde{\zeta }_0+\left(\tau _1^2+\tau _2^2\right) \tilde{\zeta }_1\right)+3 \left(\tau _1^2+\tau _2^2\right){}^2 \zeta ^1\right)-\left(\tau _1^2+\tau _2^2\right){}^3 \zeta ^0\right)}{\tau _2^3}
\right)
\\
F_{q_0}&=
\left(
2 \zeta ^1-2 \tau _1 \zeta ^0,-2 \tau _2 \zeta ^0,2  (\zeta ^0)^2+\frac{e^{2 U}}{\tau _2^3},2 \zeta ^0 \zeta^1+\frac{\tau _1 e^{2 U}}{\tau _2^3},-\tilde{\zeta }_0 \zeta ^0-\tilde{\zeta }_1 \zeta ^1-\sigma +\frac{\tau _1^3 e^{2 U}}{\tau _2^3},
\right. \\ & \left. 
-6 (\zeta ^1)^2-\frac{3 \tau _1^2 e^{2 U}}{\tau _2^3},
\zeta ^0,-\tilde{\zeta }_1 \zeta ^0 \zeta^1-\frac{e^{2 U} \left(\tau _1 \left(\tilde{\zeta }_1-\tau _1^2 \zeta ^0+3 \tau _1 \zeta ^1\right)+\tilde{\zeta }_0\right)}{\tau _2^3}-\tilde{\zeta }_0  (\zeta ^ 0)^2+\sigma  \zeta ^0-2 (\zeta ^1 )^3
\right)
\\
F_{p^1} &=
\frac{1}{3\sqrt{3} \tau_2^3}
\left(
2 \tau _2^3 \left(-\tau _1 \tilde{\zeta }_1+3 \tilde{\zeta }_0-6 \tau _1^2 \zeta ^1+6 \tau _2^2 \zeta ^1\right),-2 \tau _2^4 \left(\tilde{\zeta }_1+12 \tau _1 \zeta ^1\right),9 \left(2 \tau _2^3  (\zeta ^1)^2+\tau _1^2 e^{2 U}\right),
\right. \\ & \left. 
\tau _2^3 \left(3 \tilde{\zeta }_0 \zeta ^0-5 \tilde{\zeta }_1 \zeta ^1-3 \sigma \right)+\left(9 \tau _1^3+6 \tau _2^2 \tau _1\right) e^{2 U},9 \tau _1 \left(\tau _1^2+\tau _2^2\right){}^2 e^{2 U}-6 \tau _2^3 \tilde{\zeta }_0 \tilde{\zeta }_1,
\right. \\ & \left. 
-2 \tau _2^3 \left(18 \tilde{\zeta }_0 \zeta ^1+\tilde{\zeta }_1^2\right)-9 \left(\tau _1^2+\tau _2^2\right) \left(3 \tau _1^2+\tau _2^2\right) e^{2 U},-3 \tau _2^3 \tilde{\zeta }_1,
\right. \\ & \left. 
3 e^{2 U} \left(3 \left(\tau _1 \left(\left(\tau _1^2+\tau _2^2\right){}^2 \zeta ^0-\tau _1 \tilde{\zeta }_0\right)-\left(\tau _1^2+\tau _2^2\right) \left(3 \tau _1^2+\tau _2^2\right) \zeta ^1\right)-\tau _1 \left(3 \tau _1^2+2 \tau _2^2\right) \tilde{\zeta }_1\right)
\right. \\ & \left. +
\tau _2^3 \left(-3 \tilde{\zeta }_1 \left(\tilde{\zeta }_0 \zeta ^0+\sigma \right)-18 \tilde{\zeta }_0  (\zeta ^ 1)^2+\tilde{\zeta }_1^2 \zeta ^1\right)
\right)
\\
F_{q_1} &=
\frac{1}{3\sqrt{3} \tau_2^3}
\left(
2 \tau _2^3 \left(-2 \tilde{\zeta }_1-3 \tau _1^2 \zeta ^0+3 \tau _2^2 \zeta ^0-3 \tau _1 \zeta ^1\right),-6 \tau _2^4 \left(2 \tau _1 \zeta ^0+\zeta ^1\right),9 \left(2 \tau _2^3 \zeta ^ 0 \zeta^1+\tau _1 e^{2 U}\right),
\right. \\ & \left. 
2 \tau _2^3 \left(3 ( \zeta ^ 1)^2 -2 \tilde{\zeta }_1 \zeta ^0\right)+3 \left(3 \tau _1^2+\tau _2^2\right) e^{2 U},2 \tau _2^3 \tilde{\zeta }_1^2+9 \tau _1^2 \left(\tau _1^2+\tau _2^2\right) e^{2 U},
\right. \\ & \left. 
-3 \tau _2^3 \left(3 \tilde{\zeta }_0 \zeta ^0-5 \tilde{\zeta }_1 \zeta ^1+3 \sigma \right)-9 \tau _1 \left(3 \tau _1^2+2 \tau _2^2\right) e^{2 U},9 \tau _2^3 \zeta ^1,
\right. \\ & \left. 
\tau _2^3 \left(-9 \tilde{\zeta }_0 \zeta ^0 \zeta^1+2 \tilde{\zeta }_1^2 \zeta ^0+3 \tilde{\zeta }_1 (\zeta ^1)^2+9 \sigma  \zeta ^1\right)
\right. \\ & \left. 
-3 e^{2 U} \left(3 \tau _1^2 \left(\tilde{\zeta }_1-\tau _2^2 \zeta ^0\right)+3 \tau _1 \left(\tilde{\zeta }_0+2 \tau _2^2 \zeta ^1\right)+\tau _2^2 \tilde{\zeta }_1-3 \tau _1^4 \zeta ^0+9 \tau _1^3 \zeta ^1\right)
\right)
\end{aligned}
\end{equation}
The last Killing vector $F$ is more complicated and can be derived using the commutation relation $[F_{p^I},F_{q_J}]= 2 \delta ^I_J F$.

The corresponding Killing prepotentials

\begin{equation}
\begin{aligned}
&\Ft_{p^0} =
\left(
\begin{tabular}{c}
$\frac{e^{-U} \left(\tau _2 \left(9 \tau _1^2 \left(\tilde{\zeta }_0 \zeta ^0+\tilde{\zeta }_1 \zeta ^1-\sigma \right)-3 \tau _2^2 \left(\tilde{\zeta }_0 \zeta ^0+\tilde{\zeta }_1 \zeta ^1-\sigma \right)+4 \tau _1 \tilde{\zeta }_1^2+6 \tilde{\zeta }_0 \tilde{\zeta }_1\right)+3 \tau _1 \left(\tau _1^2-3 \tau _2^2\right) e^{2 U}\right)}{6 \sqrt{2} \tau _2^{3/2}}$ 		\\
$\frac{e^{-U} \left(-3 \tau _1^3 \left(\tilde{\zeta }_0 \zeta ^0+\tilde{\zeta }_1 \zeta ^1-\sigma \right)+\tau _1 \left(9 \tau _2^2 \left(\tilde{\zeta }_0 \zeta ^0+\tilde{\zeta }_1 \zeta ^1-\sigma \right)-6 \tilde{\zeta }_0 \tilde{\zeta }_1\right)-2 \tau _1^2 \tilde{\zeta }_1^2+2 \tau _2^2 \tilde{\zeta }_1^2-6 \tilde{\zeta }_0^2+\left(9 \tau _1^2 \tau _2-3 \tau _2^3\right) e^{2 U}\right)}{6 \sqrt{2} \tau _2^{3/2}}$	 	\\
$\frac{1}{54} e^{-2 U} \left(-27 \tilde{\zeta }_0^2 \zeta ^0-27 \tilde{\zeta }_0 \tilde{\zeta }_1 \zeta ^1+27 \sigma  \tilde{\zeta }_0+2 \tilde{\zeta }_1^3\right)-\frac{3 \tau _1 \tilde{\zeta }_0+\left(\tau _1^2+\tau _2^2\right) \tilde{\zeta }_1}{2 \tau _2}$
\end{tabular}	
\right)
\\ \\
&\Ft_{q_0} =
\left(
\begin{tabular}{c}
$-\frac{e^{-U} \left(2 \tau _2 \left(6 \tau _1 \zeta ^0 \zeta ^1-3 \tau _1^2 \zeta ^{2 0}+\tau _2^2 \zeta ^{2 0}-3 (\zeta ^1)^2\right)+e^{2 U}\right)}{2 \sqrt{2} \tau _2^{3/2}}$		\\
$-\frac{e^{-U} \left(\tilde{\zeta }_0 \zeta ^0+\tilde{\zeta }_1 \zeta ^1-6 \tau _1^2 \zeta ^0 \zeta ^1+6 \tau _2^2 \zeta ^0 \zeta ^1+6 \tau _1 \left((\zeta ^1)^2-\tau _2^2 \zeta ^{2 0}\right)+2 \tau _1^3 \zeta ^{2 0}+\sigma \right)}{2 \sqrt{2} \tau _2^{3/2}}$	 	\\
$-\frac{1}{2} e^{-2 U} \left(\tilde{\zeta }_1 \zeta ^0 \zeta ^1+\zeta ^0 \left(\tilde{\zeta }_0 \zeta ^0+\sigma \right)+2 \zeta ^{3 1}\right)-\frac{3 \left(\tau _1 \zeta ^0-\zeta ^1\right)}{2 \tau _2}$
\end{tabular}	
\right)
\\ \\
&\Ft_{p^1} =
\left(
\begin{tabular}{c}
$\frac{e^{-U} \left(2 \tau _2 \left(\tau _1 \left(-9 \tilde{\zeta }_0 \zeta ^0+15 \tilde{\zeta }_1 \zeta ^1+9 \sigma \right)+18 \tilde{\zeta }_0 \zeta ^1+\tilde{\zeta }_1^2+27 \tau _1^2 (\zeta ^1)^2-9 \tau _2^2 (\zeta ^1)^2\right)-9 \left(\tau _1-\tau _2\right) \left(\tau _1+\tau _2\right) e^{2 U}\right)}{6 \sqrt{6} \tau _2^{3/2}}$		\\
$\frac{e^{-U} \left(3 \tau _1^2 \left(3 \tilde{\zeta }_0 \zeta ^0-5 \tilde{\zeta }_1 \zeta ^1-3 \sigma \right)-9 \tau _2^2 \tilde{\zeta }_0 \zeta ^0+15 \tau _2^2 \tilde{\zeta }_1 \zeta ^1+\tau _1 \left(-36 \tilde{\zeta }_0 \zeta ^1-2 \tilde{\zeta }_1^2+54 \tau _2^2 (\zeta ^1)^2\right)-6 \tilde{\zeta }_0 \tilde{\zeta }_1-18 \tau _1^3 (\zeta ^1)^2+9 \sigma  \tau _2^2-18 \tau _1 \tau _2 e^{2 U}\right)}{6 \sqrt{6} \tau _2^{3/2}}$	 	\\
$\frac{e^{-2 U} \left(\tau _2 \left(3 \tilde{\zeta }_1 \left(\sigma -\tilde{\zeta }_0 \zeta ^0\right)-18 \tilde{\zeta }_0 (\zeta ^1)^2+\tilde{\zeta }_1^2 \zeta ^1\right)-3 e^{2 U} \left(\tau _1 \tilde{\zeta }_1-3 \tilde{\zeta }_0+6 \left(\tau _1^2+\tau _2^2\right) \zeta ^1\right)\right)}{6 \sqrt{3} \tau _2}$
\end{tabular}	
\right)\\ \\
&\Ft_{q_1} =
\left(
\begin{tabular}{c}
$\frac{e^{-U} \left(\tau _2 \left(8 \tau _1 \tilde{\zeta }_1 \zeta ^0+3 \tilde{\zeta }_0 \zeta ^0-5 \tilde{\zeta }_1 \zeta ^1+18 \tau _1^2 \zeta ^0 \zeta ^1-6 \tau _2^2 \zeta ^0 \zeta ^1-12 \tau _1 (\zeta ^1)^2+3 \sigma \right)-3 \tau _1 e^{2 U}\right)}{2 \sqrt{6} \tau _2^{3/2}}$		\\
$\frac{e^{-U} \left(6 \tau _1^2 \left(3 (\zeta ^1)^2-2 \tilde{\zeta }_1 \zeta ^0\right)+2 \left(\tau _2^2 \left(6 \tilde{\zeta }_1 \zeta ^0-9 (\zeta ^1)^2\right)+\tilde{\zeta }_1^2\right)+3 \tau _1 \left(-3 \tilde{\zeta }_0 \zeta ^0+5 \tilde{\zeta }_1 \zeta ^1+18 \tau _2^2 \zeta ^0 \zeta ^1-3 \sigma \right)-18 \tau _1^3 \zeta ^0 \zeta ^1-9 \tau _2 e^{2 U}\right)}{6 \sqrt{6} \tau _2^{3/2}}$	 	\\
$\frac{e^{-2 U} \left(\tau _2 \left(-9 \tilde{\zeta }_0 \zeta ^0 \zeta ^1+2 \tilde{\zeta }_1^2 \zeta ^0+3 \tilde{\zeta }_1 (\zeta ^1)^2-9 \sigma  \zeta ^1\right)-3 e^{2 U} \left(2 \tilde{\zeta }_1+3 \left(\tau _1^2+\tau _2^2\right) \zeta ^0+3 \tau _1 \zeta ^1\right)\right)}{6 \sqrt{3} \tau _2}$
\end{tabular}	
\right)
\end{aligned}
\end{equation}
$\Ft$ is too lengthly to be displayed.

\bibliographystyle{ssg}


\bibliography{N1Supergravity}

\end{document}